\documentclass[final]{agujournal2019}
\usepackage{booktabs}
\usepackage{amsmath}
\usepackage{amsfonts} 

\nolinenumbers
\renewcommand{\linenumbers}[1][]{}
\makeatletter
\let\@oddhead\@empty
\let\@evenhead\@empty
\renewcommand{\abstract}{\par\addvspace{18pt}\noindent{\bfseries Abstract}\vskip-\parskip
\global\c@footnote=0\relax\noindent\ignorespaces}
\makeatother

\journalname{Water Resources Research}

\begin{document}

\title{On the Adversarial Robustness of Hydrological Models}

\authors{Yang Yang\affil{1,2}, Joseph Janssen\affil{3,4}, Hoshin Gupta\affil{5}, and Ting Fong May Chui\affil{2}}

\affiliation{1}{School for the Environment, University of Massachusetts Boston, Boston, MA, USA}
\affiliation{2}{Department of Civil Engineering, The University of Hong Kong, Hong Kong SAR, China}
\affiliation{3}{Department of Geophysics, Stanford University, Palo Alto, CA, USA}
\affiliation{4}{Department of Earth, Ocean, and Atmospheric Sciences, University of British Columbia, Vancouver, BC, Canada}
\affiliation{5}{Department of Hydrology and Atmospheric Science, The University of Arizona, Tucson, AZ, USA}

\correspondingauthor{Ting Fong May Chui}{maychui@hku.hk}

\begin{abstract}
  The evaluation of hydrological models is essential for both model selection and reliability assessment. However, comparing predictions with observations alone is insufficient to fully characterize model behavior. This limitation is particularly relevant to deep learning models because of their complex internal structures. In risk-averse operational settings, water managers require models whose reliability and failure modes have been carefully evaluated because unstable predictions can put critical infrastructure at risk. Motivated by the need to assess model reliability before operational deployment, we introduce and explore adversarial robustness analysis for hydrological modeling and evaluate whether small, targeted perturbations to meteorological forcings induce substantial changes in simulated discharge. We compare a physical-conceptual model (HBV) with a deep learning-based model (LSTM) across 1,347 German catchments using perturbations of varying magnitudes generated by the fast gradient sign method (FGSM). We find that FGSM perturbations generally reduce KGE and increase MSE. However, widespread catastrophic failures under the specific FGSM perturbations are not observed, and LSTM generally exhibits less performance degradation than HBV. In the examined catchments, changes in the predicted hydrographs and time-averaged internal model states often vary approximately linearly with perturbation magnitude, providing a compact summary of how errors grow under such perturbations. Similar patterns are observed along random perturbation directions in selected catchments, suggesting that, in those catchments, small input changes produce approximately proportional changes in model output. Overall, these findings characterize local predictive stability under controlled stress tests and motivate future studies of model robustness in real-world operational settings.
\end{abstract}

\section{Introduction}

\subsection{Deep learning-based hydrological models must be provably safe}

The past decade has seen rapid growth in research investigating the use of deep learning (DL) for modeling hydrological systems \cite{tyralis2019brief,nearing2021role,shen2021applications,xu2021machine}. However, the application of DL to operational and high-impact settings, such as operational flood forecasting by government agencies, remains limited \cite{demargne2014science,luo2015clever,luo2018operational,moore1990real,gao2023artificial}. Although this disconnect likely has many causes, key barriers include the limited interpretability of many DL models and operator mistrust of ``black-box'' predictions \cite{gao2023artificial}.

DL models, particularly long short-term memory networks (LSTMs) \cite{hochreiter1997long}, have repeatedly been shown to outperform process-based conceptual models in both gauged and ungauged streamflow prediction settings \cite{lees2021benchmarking,kratzert2019towards,sabzipour2023comparing,kratzert2019toward}. For example, \citeA{lees2021benchmarking} showed that LSTMs outperformed the process-based Sacramento model for gauged streamflow prediction in the United Kingdom, while \citeA{kratzert2019toward} found that LSTMs evaluated in ungauged settings outperformed the Sacramento model evaluated in gauged settings on the CAMELS dataset \cite{addor2017camels}. Despite these successes, many practitioners remain hesitant to rely on DL-derived predictions in operational settings because their reliability under operational perturbations and distribution shifts has not been fully established \cite{Kim2026Deep,Courvoisier2025Assessing}: small, hard-to-detect input errors or distribution shifts may trigger large and unexpected departures from physically plausible behavior, and such failure modes can be difficult to anticipate. Physically-based models, in contrast, have a long history of operational use, and their structured internals are generally easier to inspect when diagnostics are needed. This gap reflects a mismatch between evaluation on historical observations and deployment under future operational conditions where data are sparse or unavailable. Chance correlations in the training data, missing rare events, and shifts driven by climate or land-use change can all create regimes in which models behave unexpectedly, even when standard test-set performance appears strong \cite{guikema2020artificial}.

For operational use, users need credible assurances that a model will behave safely across the full range of inputs within its intended operational domain, including inputs not anticipated in advance. Ideally, model developers should be able to guarantee such behavior throughout a clearly specified domain. Although establishing this level of safety is a long-term goal, an important step is to evaluate model reliability under specified input shifts and changing environmental conditions. This need motivates the present study.

\subsection{Are conventional process-based models provably safe?}

Concerns about the trustworthiness of DL models in hydrology stem partly from their black-box nature and high complexity. Conceptual hydrological models and simpler statistical models, such as linear regression, are generally easier to analyze because their structures, parameter effects, and response mechanisms are more explicit. This transparency allows users to anticipate how input changes propagate, diagnose likely failure modes, and form expectations about how the models may fail. Consider a linear model with coefficients $\beta_1, \beta_2, \ldots, \beta_p$ that predicts streamflow from lagged precipitation, analogous to a discrete unit-hydrograph formulation. A one-unit increase in the covariate $X_j$ (precipitation from $j$ days earlier) changes the predicted streamflow by $\beta_j$, holding all other covariates fixed. Likewise, in a physical-conceptual model that enforces mass balance, input changes propagate through modeled water stores and fluxes in a structured and inspectable manner. Even so, these properties do not guarantee safety, as structural, parameter, and forcing errors can still produce unreliable predictions. In contrast, DL models without explicit physical constraints provide fewer structural grounds for anticipating their responses and may produce substantially different predictions for similar inputs \cite{goodfellow2014explaining}. These catastrophic failures may arise when DL models rely on spurious relationships that do not remain stable outside the training data. Exposing the spurious and highly nonlinear model components that contribute to such failures remains difficult. Nonetheless, the issue is of major importance and has been explored widely in the fields of adversarial robustness testing and adversarial example detection \cite{carlini2019evaluating,zimmermann2022increasing,wang2019improving,ye2019adversarial}.

\subsection{Adversarial robustness testing as a way to gauge model reliability}

\begin{figure}
  \centering
  \includegraphics[width=\textwidth]
  {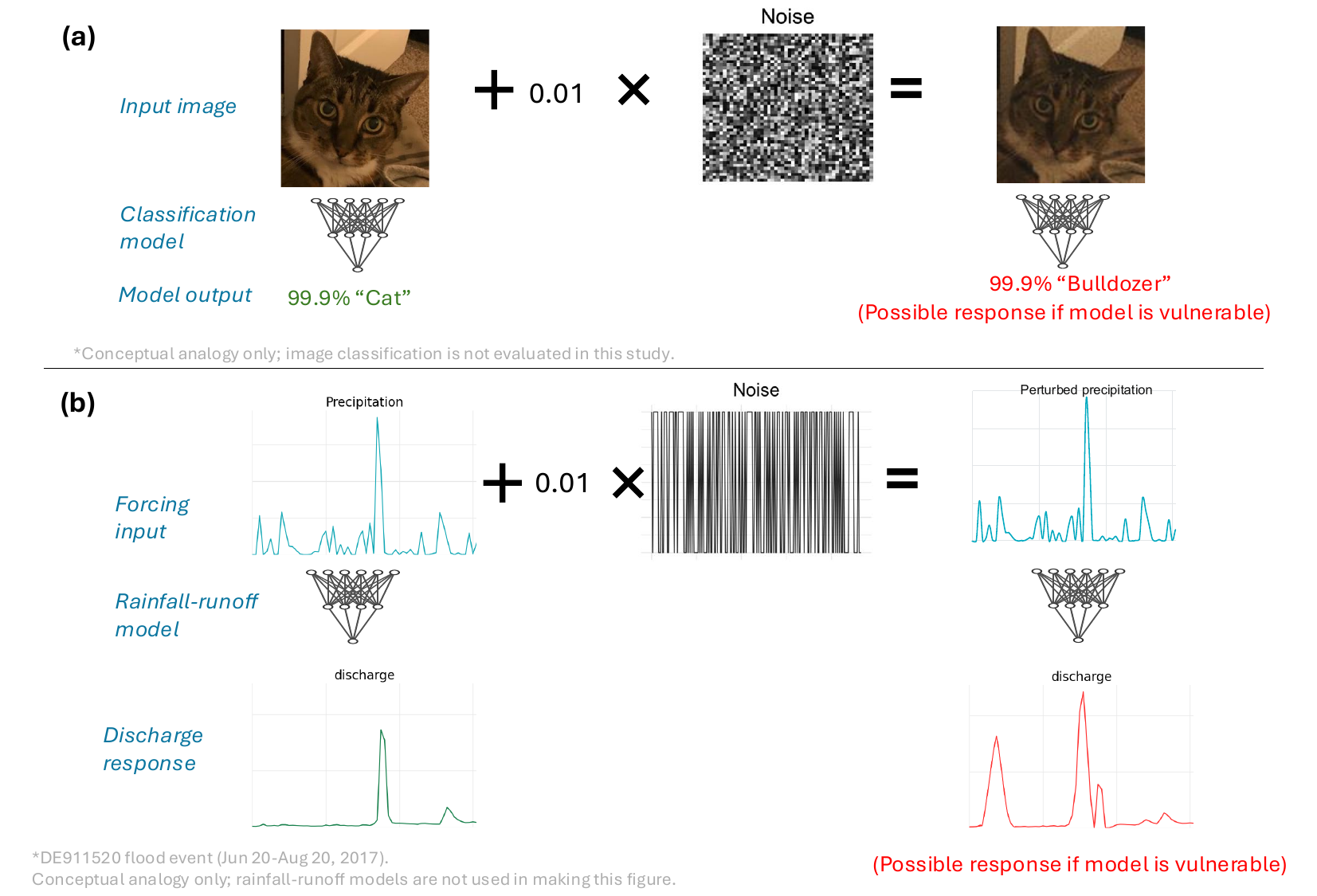}
  \caption{Conceptual illustration of adversarial vulnerability in (a) image classification and (b) rainfall-runoff modeling. Both panels are schematic and do not report experimental results. Panel (a) uses photographs of a cat as an image-classification analogy; no image-classification model was evaluated. The cat image is from Joseph Janssen. Panel (b) uses precipitation forcing from CAMELS-DE catchment DE911520 from June 20 to August 20, 2017, to illustrate the analogous rainfall-runoff setting; the discharge responses are schematic, and no rainfall-runoff model was evaluated to produce this panel. This figure is inspired by the adversarial-example illustration of \citeA{goodfellow2014explaining}.}
  \label{fig:vulnerability_illustration}
\end{figure}

An \textit{adversarial example} refers to a modified input constructed by adding an intentionally designed, imperceptible perturbation to a reference input such that the resulting input causes a specified model to produce an incorrect or altered output \cite{wiyatno2019adversarial,szegedy2013intriguing}. The modification added to the original input is called an \textit{adversarial perturbation}, and the process of constructing such a perturbation for a specific model is called an \textit{adversarial attack}. A model is \textit{adversarially robust with respect to a specified threat model} if perturbations within the allowed set and budget do not cause performance degradation beyond a predefined failure criterion. Adversarial robustness testing has been an important tool within DL since the early 2010s \cite{szegedy2013intriguing,huang2011adversarial,biggio2013evasion}, although it has been studied in the context of email spam classification with simpler machine learning and statistical models since the mid-2000s \cite{dalvi2004adversarial,barreno2006can}. More recently, adversarial robustness testing has been widely applied to image classification; a well-known example is a stop sign being misclassified as a different road sign after imperceptible modifications to the input image \cite{papernot2017practical}. Figure~\ref{fig:vulnerability_illustration}a illustrates this vulnerability schematically: a nearly imperceptible, deliberately designed perturbation causes an image classifier to change its prediction dramatically. The same idea extends to time-series regression by perturbing multivariate input time series in directions that maximize prediction error. Figure~\ref{fig:vulnerability_illustration}b shows the hydrological analogue, in which a small targeted perturbation to meteorological forcings could produce a disproportionate change in the predicted hydrograph and affect risk-sensitive decisions.

Although research on adversarial robustness in time-series prediction remains limited, existing studies demonstrate that such attacks can reveal substantial vulnerabilities \cite{kong2023adversarial,wang2023wasserstein}. For the Google stock data set, \citeA{mode2020adversarial} found that an adversarial attack increased the LSTM root mean square error (RMSE) by 27.2\%. Similarly, on their weather-prediction data set, \citeA{dera2023trustworthy} found approximately seven- to fourteen-fold RMSE increases across the evaluated LSTM variants under the strongest perturbations they tested. By contrast, \citeA{galib2023susceptibility} found that all evaluated models were vulnerable to adversarial attacks, with GRU and RNN models particularly susceptible, while LSTM and GRU showed better recovery after adversarial defense. Although other DL architectures used in hydrology may also be vulnerable to adversarial attacks, we focus on LSTMs because of their strong predictive performance and widespread use \cite{kratzert2018rainfall,shen2021applications,liu2025rnns}.

At first glance, adversarial robustness testing may seem similar to sensitivity analysis and metamorphic testing, which have been used to evaluate hydrological models \cite{mccuen1973role,gan2014comprehensive,reichert2024metamorphic,yang2021reliability}. All three approaches examine how model outputs respond to modified inputs, but they serve distinct purposes. Metamorphic testing evaluates whether outputs satisfy predefined relationships under controlled input changes. Conventional sensitivity analysis uses prescribed or sampled perturbations to characterize response patterns, whereas adversarial robustness testing selects a model-specific perturbation direction designed to maximize prediction loss within a specified perturbation budget. We therefore position adversarial robustness analysis as a complementary approximate local worst-case stress test, rather than as a standalone model-selection criterion or a replacement for conventional performance evaluation. It can reveal vulnerable forcing directions and potential catastrophic failure modes, although it cannot establish whether a model's response is physically correct when the corresponding real-world response is unknown.

Note that performance degradation under a targeted perturbation reflects predictive stability, not physical plausibility. A model's response depends on its internal state and configuration, whereas assessing its physical plausibility requires knowing how the real catchment would respond. Because that response is (usually) unknown for our constructed perturbations, a large local change may still be physically plausible and cannot by itself be taken as evidence of model failure. We therefore interpret adversarial robustness only as stability under the specified perturbation and do not treat differences in robustness as evidence that one model is generally superior.

\subsection{Why should hydrologists be concerned about adversarial robustness?}

Adversarial attacks using carefully designed input perturbations to steer water cyber-physical systems toward harmful outcomes are rarely discussed in hydrology. Yet reported cyber incidents targeting water infrastructure are increasing, and others may remain undetected or undisclosed \cite{tuptuk2021systematic}. As critical infrastructure, water facilities are attractive targets and subject to mandated protection in many countries \cite{hassanzadeh2020review}. Adversaries could subtly alter sensor readings at multiple observation points \cite{tuptuk2021systematic}. Because such manipulations can be difficult to detect, hydrologists may be required to develop models that are not overly sensitive to these perturbations. In one incident at a New York dam, an attacker accessed operational information, including system status, water levels, and temperature, rather than directly altering control actions \cite{hassanzadeh2020review}. Such access could enable later sensor manipulation, leading to unreliable predictions and poor operational decisions.

While the implications for water managers may be most immediate, evaluating adversarial robustness also informs hydrological model development and evaluation. First, adversarially robust models tend to align better with human intuitions about process behavior \cite{ortiz2021optimism}. If hydrological LSTMs demonstrate such robustness, their local responses to input changes may be more consistent with hydrologists' expectations.

Second, appropriately tuned adversarial training has been reported to improve accuracy under common image corruptions, a specific form of synthetic distribution shift \cite{ortiz2021optimism,kireev2022effectiveness}. Regional LSTMs can perform well in held-out catchments within the same large-sample domain \cite{kratzert2019toward}, yet their performance may decline when extrapolated further in space \cite{feng2023suitability}. Thus, historical or in-domain skill does not guarantee reliable behavior under climate or land-use change or during extreme events because learned relationships may extrapolate implausibly. For example, although their study did not examine LSTMs, \citeA{yang2021reliability} found that some machine learning models predicted lower flood magnitudes when precipitation alone was increased, illustrating counterintuitive responses to controlled input changes. Evaluating adversarial robustness therefore complements conventional performance evaluation by probing local stability under controlled perturbations and identifying vulnerabilities that may become relevant under broader distribution shifts.

Furthermore, assessing adversarial non-robustness in commonly used data-driven hydrological models can expose hidden failure modes and inform strategies for improvement. To uncover structural limitations within hydrological models, \citeA{andreassian2009hess} argued that hydrological models should be subjected to demanding ``crash tests''. Adversarial robustness evaluation can be viewed as a computational realization of this suggestion, providing a systematic and controllable means to push models beyond their comfort zones and to probe structural weaknesses under extreme yet informative conditions. Specifically, such adversarial ``crash tests'' may indicate that models rely on non-robust, shortcut, or overly simple predictive features \cite{shah2020pitfalls,geirhos2020shortcut,jo2017measuring,ilyas2019adversarial}. Indeed, neural networks may rely on simple feature sets that correlate strongly with the target rather than on more complex predictive features, partly because the latter are harder to discover and exploit \cite{ortiz2021optimism,shah2020pitfalls}.

\subsection{Research objective}

To the best of our knowledge, the adversarial robustness of rainfall-runoff models has not been systematically investigated. We do not know whether commonly used models will fail catastrophically when subjected to small, deliberately designed perturbations. We therefore introduce this line of research in hydrological modeling through the following objectives:

\begin{enumerate}
  \item Quantify how predictive performance changes with adversarial perturbations of different magnitudes for both DL-based and process-based models across many catchments.
  \item Examine the internal mechanisms of the prediction generation process to understand why the observed responses occur in each model.
  \item Discuss what these findings imply for model assessment, including the opportunities and limitations of robustness evaluation.
\end{enumerate}

This study addresses these objectives by comparing a popular physical-conceptual hydrological model (HBV) with an LSTM-based counterpart. In Section \ref{sec:methods}, we first describe the HBV and LSTM model architectures and then specify the adversarial perturbations used in the study. In Section \ref{sec:CaseStudy}, we introduce the modeling data, explain the experimental design, and describe the robustness evaluation procedure. The results are presented in Section \ref{sec:Results}, followed by a discussion of their implications in Section \ref{sec:discussion}.

\section{Methods}
\label{sec:methods}

\subsection{Hydrological models}
\subsubsection{HBV}\label{sec:HBV}

The Hydrologiska Byråns Vattenbalansavdelning (HBV) model is a process-based, spatially lumped hydrological model widely used in catchment-scale studies. Its popularity in both research and real-world applications is due in part to its strong performance relative to other physical-conceptual models \cite{breuer2009assessing,ayzel2021openforecast,wi2024need,chahinian2006compilation,cavadias1988approximate,bardossy2022precipitation,knoben2020brief}.

\begin{figure}[htbp]
  \centering
  \includegraphics[width=0.8\textwidth]{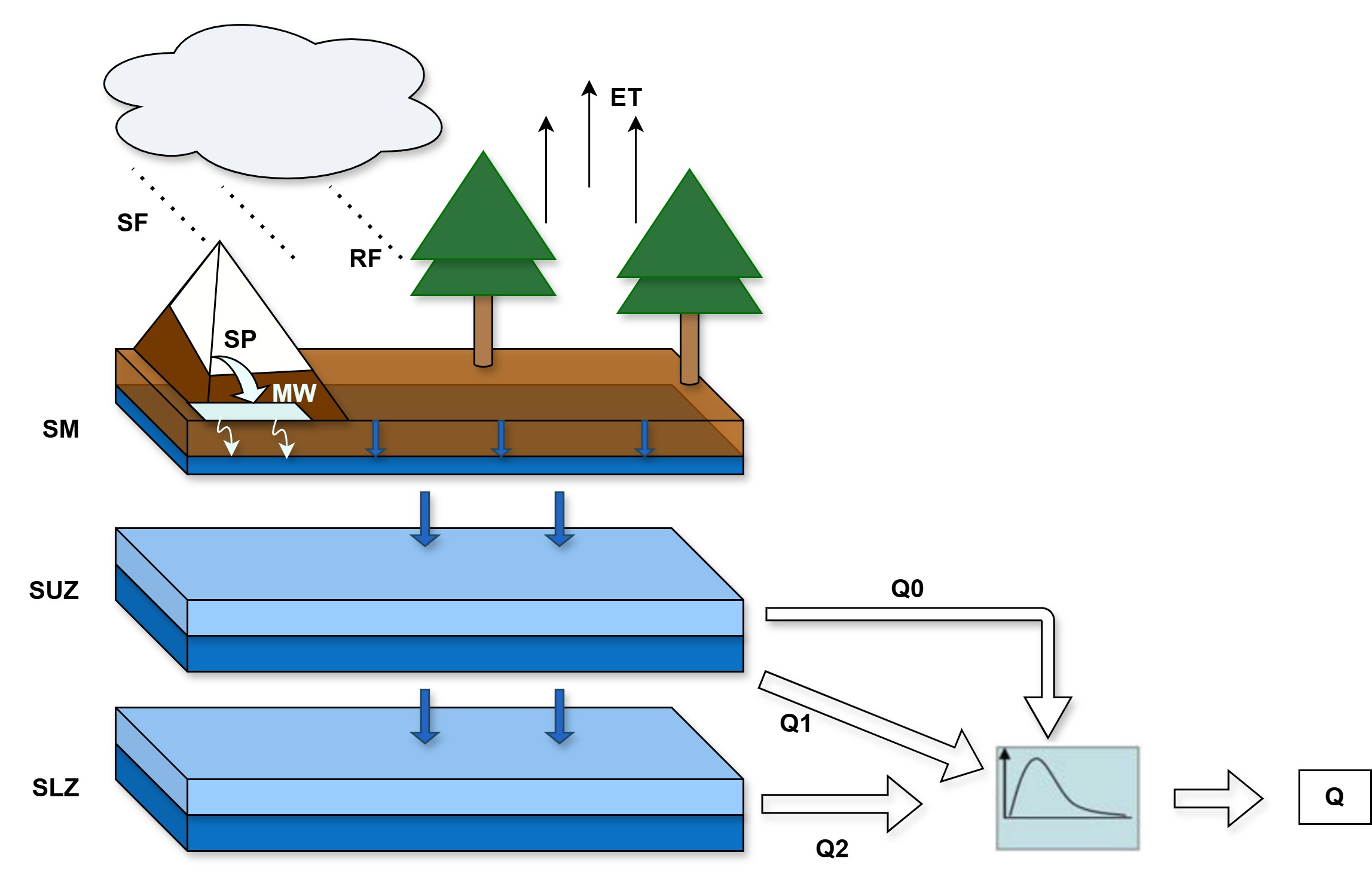}
  \caption{Structure of the HBV conceptual hydrological model. This diagram was inspired by a similar illustration presented by \protect\citeA{shrestha2008data}.
    The storages and fluxes shown include SF (snowfall), RF (rainfall), ET (evapotranspiration), SP (snowpack), MW (meltwater), SM (soil moisture), SUZ (upper-zone storage),
  SLZ (lower-zone storage), and Q (streamflow).}
  \label{figure:hbv_structure}
\end{figure}

We selected a differentiable implementation of HBV because access to gradient information relating model inputs to predictive performance enables more computationally efficient construction of model-specific adversarial perturbations. Among the many variants of HBV \cite{Jansen2021HBV}, we adapted the implementation by \citeA{AcunaEspinoza2024} from their \textit{Hy2DL} Python library \cite{KIT-Hy2DL_github}. This implementation was, in turn, based on the dPL-HBV function developed by \citeA{feng2022differentiable} and the HBV-light software developed by \citeA{seibert2005hbv}. The structure of the HBV variant used in this study is illustrated in Figure \ref{figure:hbv_structure}.

The selected HBV variant represents a catchment using five interconnected storage components: snowpack (SP), liquid water in the snowpack (MW), soil moisture (SM), upper-zone storage (SUZ), and lower-zone storage (SLZ). The model operates at a daily time step, takes precipitation ($P$), temperature ($T$), and potential evapotranspiration ($PET$) as inputs, and produces daily streamflow discharge ($Q$). This implementation does not use soil type, leaf area index, or other catchment attributes as explicit inputs. Any influence of these characteristics is represented only implicitly through the fixed, basin-specific calibrated parameters. $P$ is partitioned into rainfall and snowfall according to a temperature threshold, while the snow module simulates accumulation, melting, and refreezing. Liquid precipitation and water released from the meltwater reservoir are partitioned between soil storage and recharge according to relative soil moisture, field capacity ($FC$), and the nonlinearity parameter ($BETA$). Actual evapotranspiration is calculated by scaling $PET$ with a soil-moisture reduction factor based on $SM/(LP \times FC)$. Here, $LP$ sets the soil-moisture threshold below which evapotranspiration is reduced. The reduction factor is capped at one, and actual evapotranspiration cannot exceed the available soil moisture. Recharge and any soil-storage excess enter the SUZ, while percolation, limited by $PERC$, transfers water from SUZ to the SLZ. Fast and intermediate runoff from SUZ are governed by $K0$, $UZL$, and $K1$, whereas baseflow from SLZ is governed by $K2$. The sum of these three runoff components is then convolved with a normalized gamma-distribution unit hydrograph, where $\alpha$ and $\beta$ are the shape and scale parameters, respectively.

The HBV model has 14 parameters that must be specified for each catchment: 12 parameters for the snow, soil, evapotranspiration, and groundwater routines and two parameters for gamma-distribution routing. The implementation supports an additional evapotranspiration parameter that is not used in this study. It uses both linear and nonlinear formulations to represent catchment-scale hydrological processes. For example, potential snowmelt is calculated as $melt = CFMAX \times (T - TT)$ when $T > TT$ and is set to zero otherwise; actual snowmelt is then limited by the available snowpack. In contrast, the fraction of liquid precipitation and released meltwater assigned to recharge is a nonlinear function of relative soil moisture, $\left(SM/FC\right)^{BETA}$, capped at one. Here, $SM$ is soil moisture, $FC$ is field capacity, and $BETA$ controls the nonlinearity of the recharge fraction. More details can be found in \citeA{AcunaEspinoza2024} and \citeA{KIT-Hy2DL_github}.

\subsubsection{LSTM}
\label{sec:LSTM}

We adopt the LSTM model architecture developed by \citeA{Yang2024generative_lumped} and drive it with the same meteorological forcings used by the HBV model (i.e., $P$, $T$, and $PET$). Specifically, we train a single regional LSTM whose network weights are shared across all catchments. To represent catchment-specific behavior, we associate each catchment $c$ with a learnable $d$-dimensional embedding vector, $\mathbf{e}_c \in \mathbb{R}^d$, which is optimized jointly with the network weights during training.

At each time step $t$, $\mathbf{e}_c$ is concatenated with the meteorological inputs $P_t$, $T_t$, and $PET_t$ to form the input vector $\mathbf{x}_t = [\mathbf{e}_c^\mathsf{T},\, P_t,\, T_t,\, PET_t]^\mathsf{T} \in \mathbb{R}^{d+3}$. We do not use physical catchment attributes as input features, consistent with the HBV baseline, which also does not use such attributes as explicit inputs. Instead, each embedding $\mathbf{e}_c$ is learned from meteorological-forcing and discharge data by minimizing the discharge-prediction loss. In this respect, learning the embedding is conceptually analogous to calibrating the parameters of a process-based model from observations.

The LSTM processes the sequence $\{\mathbf{x}_t\}$ and updates its gates and internal states as follows:
\begin{linenomath*}
  \begin{align*}
    \mathbf{f}_t &= \sigma(\mathbf{W}_f \mathbf{x}_t + \mathbf{U}_f \mathbf{h}_{t-1} + \mathbf{b}_f),
    & \text{(forget gate)} \\
    \mathbf{i}_t &= \sigma(\mathbf{W}_i \mathbf{x}_t + \mathbf{U}_i \mathbf{h}_{t-1} + \mathbf{b}_i),
    & \text{(input gate)} \\
    \widetilde{\mathbf{c}}_t &= \tanh(\mathbf{W}_c \mathbf{x}_t + \mathbf{U}_c \mathbf{h}_{t-1} + \mathbf{b}_c),
    & \text{(candidate cell state)} \\
    \mathbf{o}_t &= \sigma(\mathbf{W}_o \mathbf{x}_t + \mathbf{U}_o \mathbf{h}_{t-1} + \mathbf{b}_o),
    & \text{(output gate)} \\
    \mathbf{c}_t &= \mathbf{f}_t \odot \mathbf{c}_{t-1}
    + \mathbf{i}_t \odot \widetilde{\mathbf{c}}_t,
    & \text{(cell-state update)} \\
    \mathbf{h}_t &= \mathbf{o}_t \odot \tanh(\mathbf{c}_t),
    & \text{(hidden-state update)} \\
    \widehat{Q}_t &= \operatorname{MLP}(\mathbf{h}_t),
    & \text{(discharge prediction)}
  \end{align*}
\end{linenomath*}
Here, $\sigma$ denotes the sigmoid activation function; $\odot$ denotes element-wise multiplication; $\mathbf{f}_t$, $\mathbf{i}_t$, and $\mathbf{o}_t$ are the forget, input, and output gates, respectively; $\widetilde{\mathbf{c}}_t$ is the candidate cell state; and $\mathbf{c}_t$ and $\mathbf{h}_t$ are the cell and hidden states. The matrices $\mathbf{W}_*$ and $\mathbf{U}_*$ and the bias vectors $\mathbf{b}_*$ are learnable parameters. The hidden state $\mathbf{h}_t$ is passed through a multilayer perceptron (MLP) with ReLU-activated hidden layers and a linear output layer to produce the predicted discharge $\widehat{Q}_t$. The hidden-state dimension, embedding dimension $d$, number of LSTM layers, and MLP architecture are selected through hyperparameter tuning.

\subsection{Adversarial robustness}
\label{sec:AdversarMethods}

\subsubsection{Fast Gradient Sign Method (FGSM)}
The Fast Gradient Sign Method (FGSM), introduced by \citeA{goodfellow2014explaining}, is widely used to generate adversarial examples. FGSM perturbs each input component in the direction indicated by the sign of the loss gradient. We compute the perturbed meteorological inputs as
\[
  \mathbf{X}_{\epsilon}
  = \mathbf{X}
  \mathbin{+} \epsilon\,\operatorname{sign}\!\left(
    \nabla_{\mathbf{X}} J(\theta,\mathbf{X},\mathbf{y})
  \right),
\]
where $\mathbf{X} \in \mathbb{R}^{3\times m}$ contains $m$ days of $P$, $T$, and $PET$; $\mathbf{X}_{\epsilon}$ is the perturbed input; and $\epsilon \in \mathbb{R}_{\geq 0}$ is the perturbation magnitude. Here, $\theta$ denotes the model-specific parameters held fixed during the adversarial experiments. For the LSTM, these comprise the shared network weights and catchment embedding $\mathbf{e}_c$; for HBV, they include the basin-specific calibrated parameters. The observed discharge vector is $\mathbf{y}\in\mathbb{R}^{n}$, where $n$ is the number of output days evaluated from the $m$-day input sequence. We use the mean squared error (MSE) between the predicted discharge $g(\mathbf{X},\theta)\in\mathbb{R}^{n}$ and the observed discharge $\mathbf{y}$ as the loss $J(\theta,\mathbf{X},\mathbf{y})$. The operator $\nabla_{\mathbf{X}}$ gives the partial derivative of the loss with respect to each element of $\mathbf{X}$, and $\operatorname{sign}$ is applied element-wise. Both the HBV and LSTM model implementations support automatic differentiation, enabling efficient computation of the input gradients.

In simple terms, we first run each model with the original meteorological inputs and calculate the prediction error between simulated and observed discharge. We then use backpropagation to determine the gradient of this error with respect to each meteorological input variable at every time step. Each gradient indicates the direction in which changing $P$, $T$, or $PET$ would increase the prediction error. FGSM retains only the sign of the gradient and changes the corresponding input by $+\epsilon$ or $-\epsilon$ in that direction, while an input with a zero gradient remains unchanged. Because HBV and LSTM differ in structure and typically yield different input gradients, they generally produce different perturbations for the same meteorological input sequence. For each model and catchment, we compute the gradient-sign direction once at the unperturbed input and hold it fixed while varying $\epsilon$.

We apply the perturbations directly to inputs expressed in their original physical units. Consequently, $\epsilon$ represents the same numerical perturbation magnitude in each variable's native unit rather than a common dimensionless or physically equivalent perturbation across variables. For example, $\epsilon=0.2$ changes each input with a nonzero gradient by 0.2~mm/day for $P$ and $PET$ or by 0.2~$^\circ$C for $T$ (before the non-negativity adjustment described below).

After applying FGSM, we enforce non-negativity constraints by setting negative perturbed values of $P$ and $PET$ to zero. We do not adjust the perturbed $T$ values because subzero temperatures are physically possible. The resulting input is denoted by $\mathbf{X}_{\epsilon,\mathrm{adj}}$.

\subsubsection{Quantifying model adversarial robustness}

The effect of an adversarial perturbation can be quantified by the change in an evaluation metric, such as the Kling--Gupta efficiency (KGE), Nash--Sutcliffe efficiency (NSE), or MSE, between the unperturbed and perturbed inputs:
\[
  \Delta \mathrm{Perf} = \mathrm{Perf}(g(\mathbf{X}_{\epsilon,\mathrm{adj}},\theta), \mathbf{y}) - \mathrm{Perf}(g(\mathbf{X},\theta), \mathbf{y}),
\]
where $\mathrm{Perf}(\cdot)$ denotes the evaluation metric, $g$ is the hydrological model, $\theta$ denotes the model-specific parameters held fixed during the adversarial experiments, $\mathbf{X}$ is the original input, $\mathbf{X}_{\epsilon,\mathrm{adj}}$ is the adjusted adversarial input at perturbation magnitude $\epsilon$, and $\mathbf{y}$ is the observed discharge. Thus, degradation yields negative $\Delta\mathrm{Perf}$ for KGE and NSE and positive $\Delta\mathrm{Perf}$ for MSE.

To systematically evaluate and compare model robustness across perturbation magnitudes $\epsilon$, we use the fitted slope as our reported summary. First, we fit a linear model between $\epsilon$ and $\Delta \mathrm{Perf}$ and report its slope. In this study, these regressions use all eleven evaluated perturbation magnitudes, $\epsilon\in\{0,0.05,0.10,\ldots,0.50\}$. Because $\Delta\mathrm{Perf}=0$ by definition at $\epsilon=0$, including this point ensures that the fitted relationship represents the full evaluated perturbation range, including the initial interval from $\epsilon=0$ to $\epsilon=0.05$. When a high coefficient of determination $R^2$ indicates that the relationship is adequately described by a linear model and the fitted slope has the degradation direction, a smaller absolute slope indicates greater adversarial robustness because performance degrades more slowly as the perturbation magnitude increases. This interpretation is specific to the defined adversarial test and does not imply that the model is generally less sensitive to meteorological forcings. Alternatively, one can compute the area under the $\epsilon$--$\mathrm{Perf}$ curve (AUC), which summarizes performance across the perturbation range under consideration without assuming a linear relationship. For a higher-is-better metric such as KGE or NSE, a larger AUC indicates better robust performance; for a lower-is-better metric such as MSE, a smaller AUC indicates better robust performance.

Such an AUC can be normalized by the width of the specified $\epsilon$ interval to express it as an average metric value over that interval:
\[
  \mathrm{RobustPerf}_{\mathrm{avg}} = \frac{1}{\epsilon_{\max} - \epsilon_{\min}} \int_{\epsilon_{\min}}^{\epsilon_{\max}} \mathrm{Perf}(g(\mathbf{X}_{\epsilon,\mathrm{adj}},\theta), \mathbf{y}) \, d\epsilon,
\]
where $\mathrm{Perf}(g(\mathbf{X}_{\epsilon,\mathrm{adj}},\theta), \mathbf{y})$ is the metric value at perturbation magnitude $\epsilon$. This normalized AUC represents the average robust performance (e.g., robust NSE or robust KGE) over $[\epsilon_{\min},\epsilon_{\max}]$ and can be approximated numerically from the evaluated $\epsilon$ values. Because robustness depends on the perturbation interval, direct model comparisons should use the same $\epsilon$ range.

\section{Experimental setup}
\label{sec:CaseStudy}

\subsection{Data}
\label{sec:Data}

We use the CAMELS-DE dataset \cite{loritz2024camels} (Version 1.0.0) for all numerical experiments in this study. CAMELS-DE provides daily hydro-meteorological forcings, discharge observations, and catchment attributes for 1,582 diverse catchments across Germany, with records spanning up to 70 years (1951--2020). We use data from 1981 to 2020 in this study.

Although CAMELS-DE includes pretrained LSTM models, the meteorological forcings used by these models differ from those used by HBV. For example, mean radiation is an input to the CAMELS-DE LSTMs, whereas HBV uses only $P$, $T$, and $PET$. We therefore train our own LSTM model, as described in Section \ref{sec:LSTM}, using the same input variables as HBV. Negative discharge values, which may arise from tidal effects or human influences such as water-resources management, are treated as missing. To ensure sufficient data for model training, validation, and testing, we include only catchments with at least two years (730 days) of nonmissing discharge data in each of the training (1981--2000), validation (2001--2010), and test (2011--2020) periods. After applying this criterion, 1,347 catchments remain for analysis. For all experiments, we evaluate the HBV and LSTM models using the same test-period data.

\subsection{HBV calibration}

The HBV variant used in this study uses 14 fixed, basin-specific parameters. We use the provided calibrated parameter sets from CAMELS-DE \cite{loritz2024camels} directly in our HBV simulations. These parameter sets were calibrated separately for each catchment using data from 1 October 1970 through 31 December 1999, NSE as the calibration objective, and the DiffeRential Evolution Adaptive Metropolis (DREAM) algorithm \cite{vrugt2016markov}. This differs from the final LSTM, which was trained regionally using MSE over 1981--2010. Neither fitting period overlaps the common 2011--2020 test period, so there is no test-data leakage and we treat both models as suitable for evaluating that period. However, differences in fitting data, objectives, and local versus regional estimation confound any attribution of between-model differences solely to model architecture. For our simulations, we use a 365-day warm-up period. The parameter values remain fixed throughout our analysis; no dynamic or regionally learned HBV parameters are used. The differentiable implementation is used only to calculate the gradients required for the adversarial robustness analysis and does not alter the basin-specific calibration framework. Consequently, the results characterize this conventional HBV configuration and may not represent the behavior of newer hybrid models with dynamically or regionally learned parameters.

\subsection{LSTM training}

To optimize the LSTM hyperparameters, we use Bayesian optimization implemented with the \textit{Optuna} Python package \cite{akiba2019optuna}. Each candidate configuration is trained on the training set and evaluated by its prediction error on the validation set. The optimized hyperparameters include the number of LSTM layers, the hidden-state dimension, the catchment-embedding dimension $d$, the learning rates for the catchment embeddings and LSTM network, the MLP architecture (i.e., the number of layers and neurons), and the training batch size. We conduct 200 optimization trials and use MSE as both the training loss and the validation objective.

During training, each batch contains randomly sampled 730-day (2-year) input sequences drawn from different catchments. The three meteorological forcing variables are supplied to the LSTM in their original physical units without normalization: $P$ and $PET$ are expressed in mm/day, and $T$ is expressed in $^\circ$C. The prediction target is the discharge during the final 365 days (i.e., the second year) of each sequence; the first 365 days serve as a warm-up period. During validation and testing, the complete input sequence for each catchment is processed in a single run. A 365-day warm-up period precedes the respective evaluation period, and only post-warm-up discharge predictions are included in the validation or test hydrograph and performance calculations. During hyperparameter tuning, training stops if validation performance does not improve for 20 consecutive epochs or after a maximum of 500 epochs.

After selecting the optimal hyperparameters, we retrain the LSTM using the combined training and validation data and evaluate the resulting model on the independent test set. For final training, we use two-thirds of the validation-selected epoch count to account for the larger combined dataset. The selected embedding dimension is $d=16$. At each time step, the model receives a 19-dimensional input comprising three meteorological forcings ($P$, $T$, and $PET$) concatenated with the learned catchment embedding. The LSTM output is passed through an MLP without dropout, with the architecture $254 \rightarrow 13$ (ReLU) $\rightarrow 5$ (ReLU) $\rightarrow 1$ (linear). The resulting model achieves a median test KGE of 0.833 and a median NSE of 0.869. For context, \citeA{loritz2024camels} reported a median NSE of 0.84 for their CAMELS-DE LSTMs; these NSE values are not directly comparable because the studies use different forcings, catchment-selection criteria, and data splits.

\subsection{Imperceptibility of adversarial examples and selection of perturbation magnitude}
Imperceptibility is a key requirement (or constraint) for adversarial attacks: perturbations should be small enough that the adversarial examples $\mathbf{X}_{\epsilon}$ are generally indistinguishable from the original inputs. While imperceptibility in images can be assessed visually by humans, for time series we use the published validation statistics as contextual references for the perturbation scale in the meteorological data as a reference. For example, $P$ from the HYRAS dataset (which was used to develop CAMELS-DE) has a mean absolute error (MAE) of 1--2~mm/day compared to gauge observations \cite{rauthe2013central}, and $T$ has an MAE of about 0.6$^\circ$C and a bias of 0.2$^\circ$C \cite{razafimaharo2020new}.

\begin{figure}[h!]
    \centering
    \includegraphics[width=\textwidth]{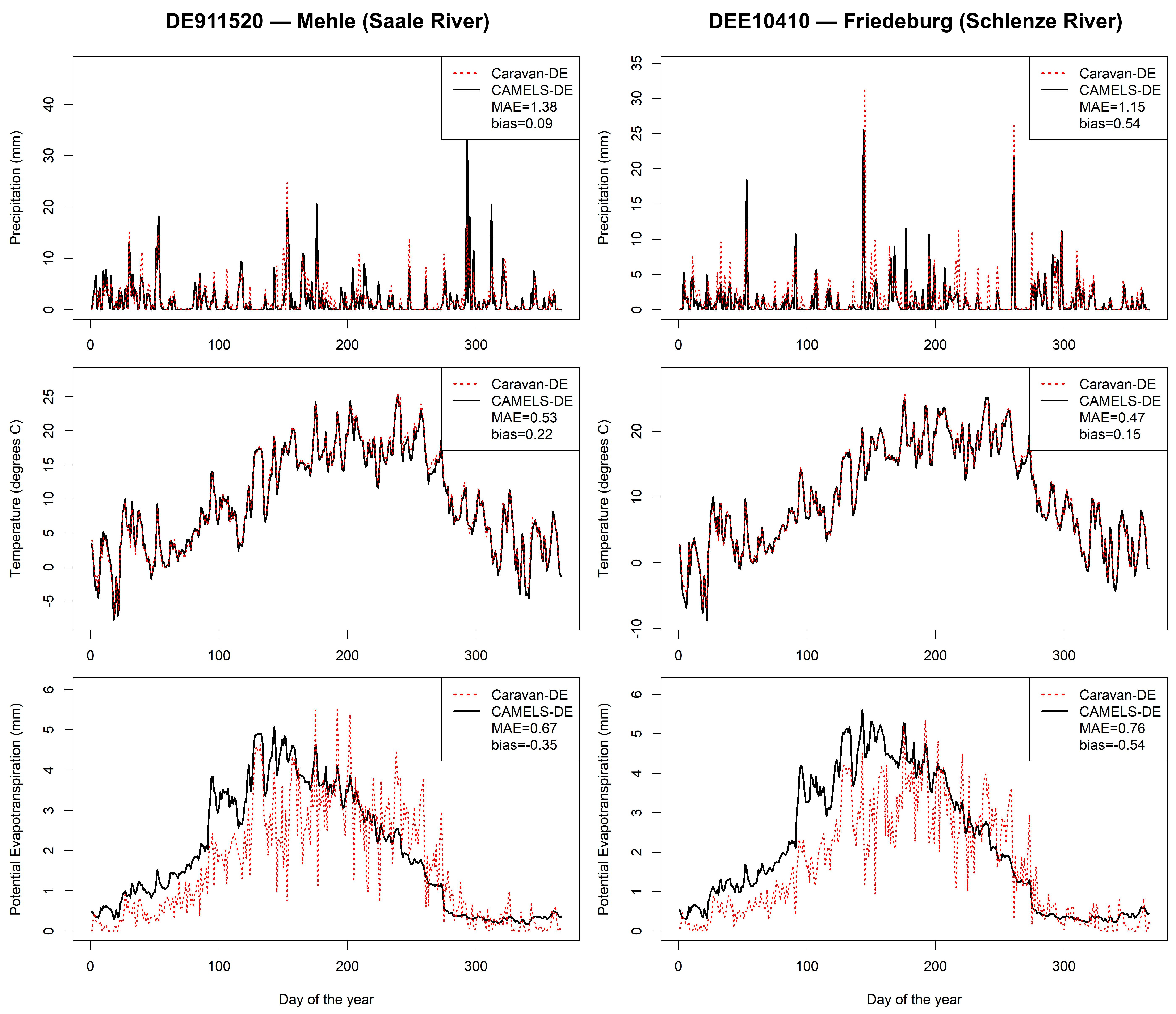}
    \caption{Comparison of CAMELS-DE (black) and Caravan-DE (red) time series for precipitation (top), temperature (middle), and potential evapotranspiration (bottom) in two catchments in 2016: a representative catchment with average precipitation (DE911520, left) and a very dry catchment in the CAMELS-DE dataset (DEE10410, right). Mean absolute error (MAE) and bias (Caravan-DE - CAMELS-DE) are also shown.}
    \label{fig:imperceptible}
  \end{figure}

To further contextualize this scale, we compared CAMELS-DE time series derived from HYRAS with those from the CAMELS-DE Caravan Extension, derived from ERA5-Land (hereafter Caravan-DE) \cite{caravande_all_versions}, for two catchments in 2016 (Figure~\ref{fig:imperceptible}). For a representative catchment DE911520 (Mehle gauge station, Saale River; 135.96 km$^2$ catchment area), whose mean precipitation is near the median across CAMELS-DE catchments, $P$ has an MAE of $1.38$~mm/day and a bias of $0.09$~mm/day, while $T$ has an MAE of 0.53$^\circ$C and a bias of $0.22$~$^\circ$C. Similar MAE and bias magnitudes are seen for $PET$. For a very dry catchment in CAMELS-DE (DEE10410; Friedeburg gauge station, Schlenze River; 108.0~km$^2$ catchment area), MAEs and biases for all variables are in the range 0.15--1.15.

  Based on these values, we choose a baseline perturbation magnitude of $\epsilon=0.2$ for our experiments, which generally falls within the range of differences observed among meteorological products. We also use a range of $\epsilon$ from 0 up to 0.5 to study model responses to different perturbation magnitudes. Visual comparisons of the original and adversarial examples are also provided to evaluate imperceptibility in Section~\ref{sec:overall_results}. Note that the CAMELS-DE--Caravan-DE comparison is used only to contextualize the magnitude of the perturbations; neither model is driven by Caravan-DE in the analysis.

  \section{Results}
  \label{sec:Results}

  \subsection{Models' overall adversarial robustness}\label{sec:overall_results}

\begin{figure}[htbp]
    \centering
    \includegraphics[width=\textwidth]{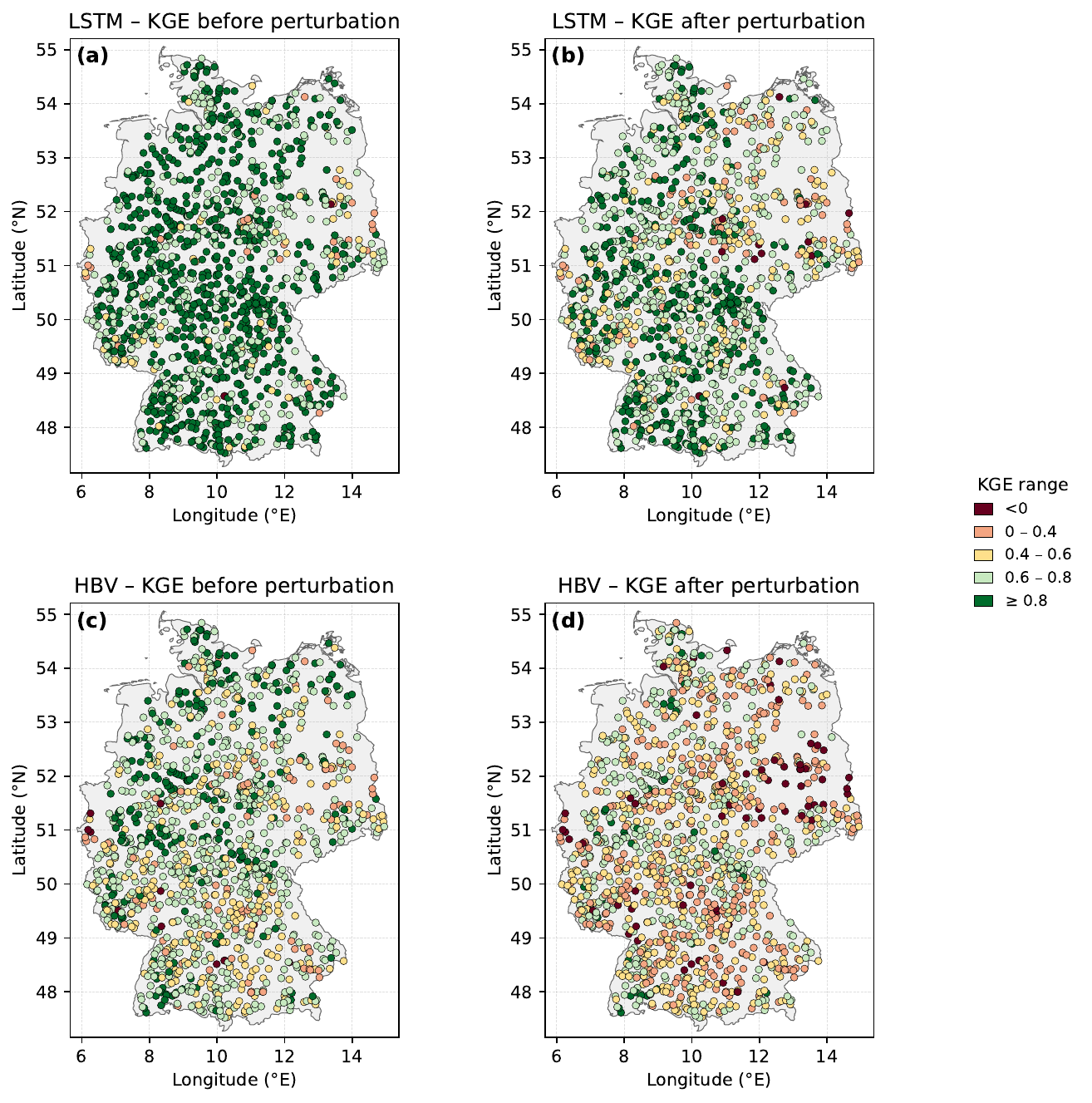}
    \caption{Maps of Kling–Gupta efficiency (KGE) across CAMELS-DE gauging stations for the LSTM and HBV models before and after FGSM adversarial perturbations of magnitude $\epsilon=0.2$. Panels show (a) LSTM before, (b) LSTM after, (c) HBV before, and (d) HBV after perturbations; point colors denote KGE categories as in the legend.}
    \label{figure:kge_maps}
  \end{figure}

\begin{figure}[htbp]
    \centering
    \includegraphics[width=\textwidth]{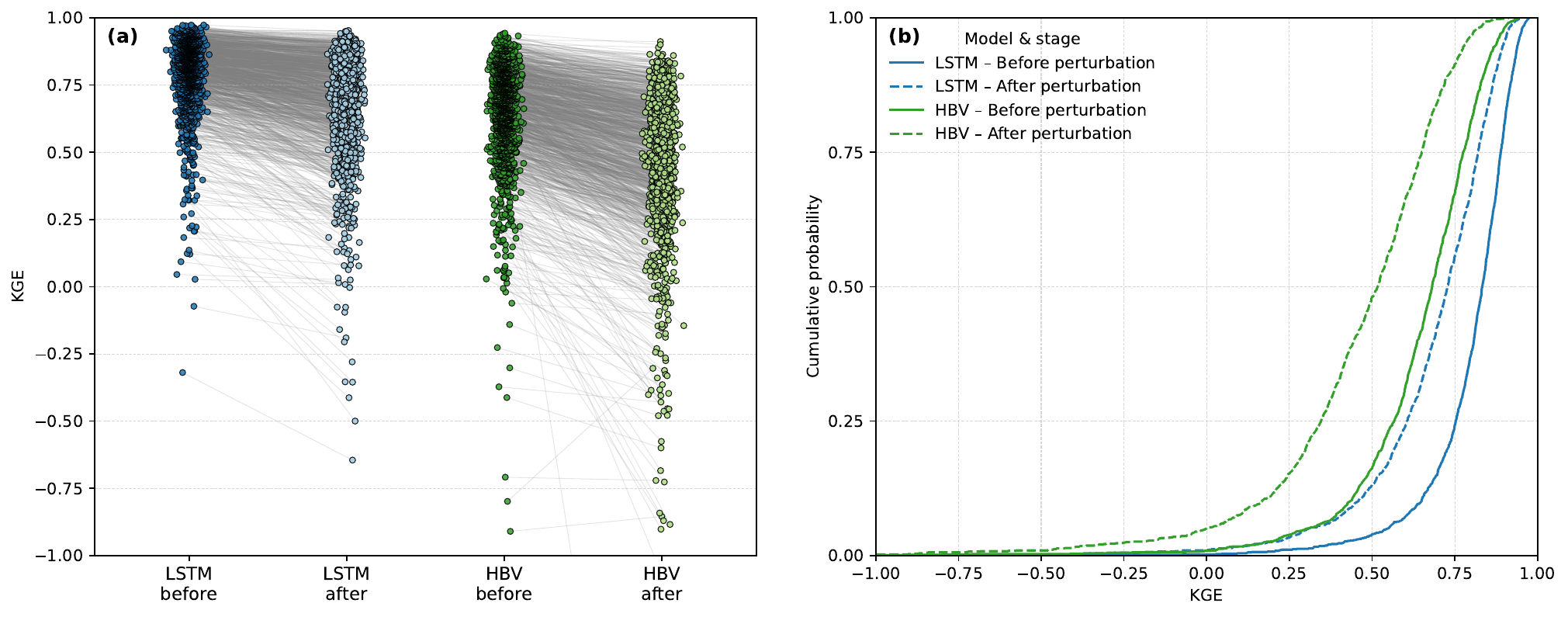}
    \caption{KGE across CAMELS-DE gauging stations before and after an FGSM perturbation ($\epsilon=0.20$) for LSTM and HBV. (a) Connected dot plot: for each catchment, a gray line links KGE before (left) to after (right) within each model (LSTM in blue, HBV in green; jitter added for visibility). (b) Empirical cumulative distribution functions of KGE, with solid lines for before-perturbation and dashed lines for after-perturbation distributions, summarizing the distributional shift. A few sites have $\mathrm{KGE}<-1$ (LSTM: 1 before/1 after; HBV: 1 before/3 after) and are outside the $[-1,1]$ axis limits.}
    \label{figure:ecdf}
  \end{figure}

  Across the 1,347 CAMELS-DE catchments, the median test KGE is 0.833 for the LSTM model and 0.681 for the HBV model (Figures \ref{figure:kge_maps}a and c). Under an FGSM perturbation of magnitude $\epsilon=0.2$, their median KGEs drop to 0.728 (LSTM; $\Delta=-0.105$) and 0.517 (HBV; $\Delta=-0.164$), respectively, with widespread deterioration visible across Germany (Figures \ref{figure:kge_maps}b and d). The connected dot plot (Figure \ref{figure:ecdf}a) shows a widespread per-catchment decline; only 5 of 1,347 catchments (0.37\%) for LSTM and 3 of 1,347 catchments (0.22\%) for HBV improve. Note that MSE increases at every catchment for both models, consistent with the use of MSE as the FGSM loss. The empirical cumulative distribution functions (ECDFs) of KGE (Figure \ref{figure:ecdf}b) shift left after the perturbations, indicating degradation across the full distribution. Despite the degradation, the perturbed LSTM still outperforms the unperturbed HBV across most quantiles, including the median (KGE is 0.728 vs. 0.681). These results show that FGSM-based perturbations can effectively degrade the performance of both LSTM and HBV models. Surprisingly, however, LSTM exhibits less performance degradation than HBV.

  Although the relationship between baseline skill and adversarial sensitivity is not consistent across models and metrics, Figure \ref{figure:ecdf}a shows that some of the largest KGE decreases occur among catchments with low unperturbed KGE. Figure \ref{figure:ecdf}b likewise shows that the before- and after-perturbation ECDFs are close toward the upper end of the distribution but more widely separated at lower cumulative probabilities. Given these observations and the fact that HBV starts with a lower KGE than LSTM, our conclusion that the LSTM-based model is more robust than HBV may be an artifact of differences in the models' initial KGE. Therefore, to examine whether this comparison persists across ranges of initial KGE, we stratify the results for both models by their initial KGE levels. As reported in Table \ref{tab:kge_table}, the same median ordering appears across all reported strata of initial KGE and across two evaluation metrics (MSE and KGE). Note that the $\Delta$KGE values in the previous paragraph are computed as differences between overall medians, whereas Table~\ref{tab:kge_table} reports the median of per-catchment drops; these summaries need not match because $\mathrm{median}(x-y)\neq \mathrm{median}(x)-\mathrm{median}(y)$.

  \begin{table}[ht]
    \centering
    \small
    \setlength{\tabcolsep}{3.5pt}
    \caption{Performance degradation under FGSM ($\epsilon=0.2$) stratified by baseline skill. For each model separately, catchments are grouped by that model's unperturbed KGE ($\mathrm{KGE}_o$). Each entry reports the median performance degradation as KGE decrease (MSE increase), calculated per catchment within the stratum. The five displayed strata cover all 1,347 catchments for each model, although their sample sizes are model-specific; the ``All'' column also summarizes all 1,347 catchments.}
    \label{tab:kge_table}
    \resizebox{\textwidth}{!}{%
      \begin{tabular}{lcccccc}
        \toprule
        & $\mathrm{KGE}_o\in[0.8,1]$ & $\mathrm{KGE}_o\in[0.7,0.8)$ & $\mathrm{KGE}_o\in[0.6,0.7)$ & $\mathrm{KGE}_o\in[0.5,0.6)$ & $\mathrm{KGE}_o\in(-\infty,0.5)$ & All \\
        \midrule
        HBV  & 0.116 (0.141) & 0.130 (0.145) & 0.163 (0.137) & 0.188 (0.128) & 0.215 (0.095) & 0.156 (0.133) \\
        LSTM & 0.067 (0.077) & 0.102 (0.074) & 0.126 (0.105) & 0.097 (0.066) & 0.123 (0.026) & 0.082 (0.075) \\
        \bottomrule
      \end{tabular}%
    }
  \end{table}

  Across all displayed baseline-skill groups, HBV exhibits greater performance degradation than LSTM. The contrast remains pronounced in the lowest displayed skill group ($\mathrm{KGE}_o<0.5$), where the median KGE decrease is approximately 1.75 times as large for HBV as for LSTM (0.215 versus 0.123).

\begin{figure}[htbp]
    \centering
    \includegraphics[width=\textwidth]{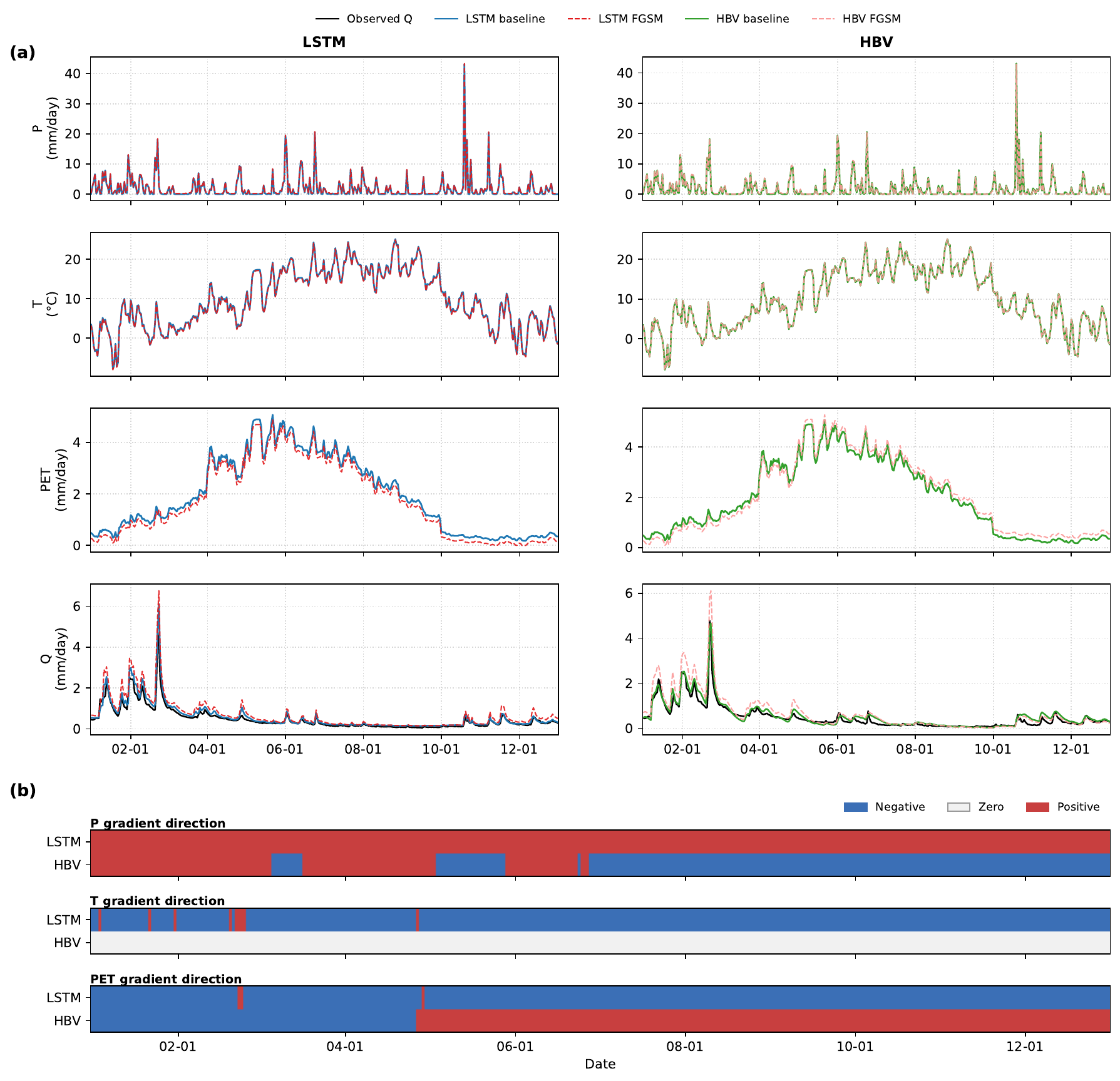}
    \caption{(a) Baseline and FGSM-perturbed meteorological forcings---precipitation ($P$), temperature ($T$), and potential evapotranspiration ($PET$)---and predicted discharge ($Q$) for the LSTM and HBV models for CAMELS-DE catchment DE911520 during 2016 at $\epsilon=0.2$. Observed discharge is shown in black. The baseline and perturbed $P$ and $T$ traces largely overlap, whereas the differences in $PET$ and $Q$ are more visible. (b) Sign of the loss gradient with respect to each meteorological input at each time step for the LSTM and HBV.}
    \label{figure:input_output_example}
  \end{figure}

  Note that the adversarial perturbations applied to the meteorological inputs are generally subtle. As an example, Figure~\ref{figure:input_output_example}a illustrates this for CAMELS-DE catchment DE911520, where the baseline and FGSM-perturbed $P$ and $T$ time series for 2016 are very similar for both the LSTM and HBV models, with relatively small differences at most time steps. Although the $PET$ traces show larger relative deviations, these $PET$ perturbations remain substantially smaller than the differences between forcings derived from different meteorological products, as shown in Figure~\ref{fig:imperceptible}. The apparent imperceptibility of the $Q$ changes also depends on discharge magnitude and flow regime: for some periods, especially low flows, the baseline and adversarial hydrographs can still look quite similar, even though predictive performance substantially degrades. This is expected because streamflow differences can reflect the accumulation of small input perturbations through internal model states. At $\epsilon=0.2$, FGSM changes each input component with a nonzero gradient by 0.2~mm/day for $P$ and $PET$ or by 0.2~$^\circ$C for $T$ before the non-negativity adjustment. For DE911520 in 2016, however, the maximum absolute daily change in predicted $Q$ reached 0.734~mm/day for LSTM and 1.444~mm/day for HBV, exceeding the 0.2~mm/day bound applied to individual $P$ and $PET$ perturbations. The normalized mean absolute perturbations in $P$, $T$, and $PET$ were (10.6\%, 2.0\%, 10.3\%) for LSTM and (7.3\%, 0.0\%, 10.3\%) for HBV, whereas the corresponding changes in $Q$ relative to the respective baseline simulations were 21.6\% and 27.3\%, respectively. The annual peak $Q$ increased by 10.6\% for LSTM and 30.9\% for HBV, while the peak date remained unchanged in both cases.

  Figure~\ref{figure:input_output_example}b shows the sign of the loss gradient with respect to each meteorological input, which determines the FGSM perturbation direction. The gradient-sign patterns differ between the LSTM and HBV. For HBV, the gradient with respect to $T$ is zero throughout 2016 despite the occurrence of snow accumulation and melt. On every melt day, potential melt exceeded the available snowpack, so the minimum operation set actual melt equal to the available snowpack and made it locally insensitive to $T$. This zero gradient results from HBV's piecewise snow-process operations and from evaluating the gradient locally at the unperturbed input; it does not imply that $T$ is irrelevant to finite changes that cross model thresholds.

\begin{figure}[htbp]
    \centering
    \includegraphics[width=\textwidth]{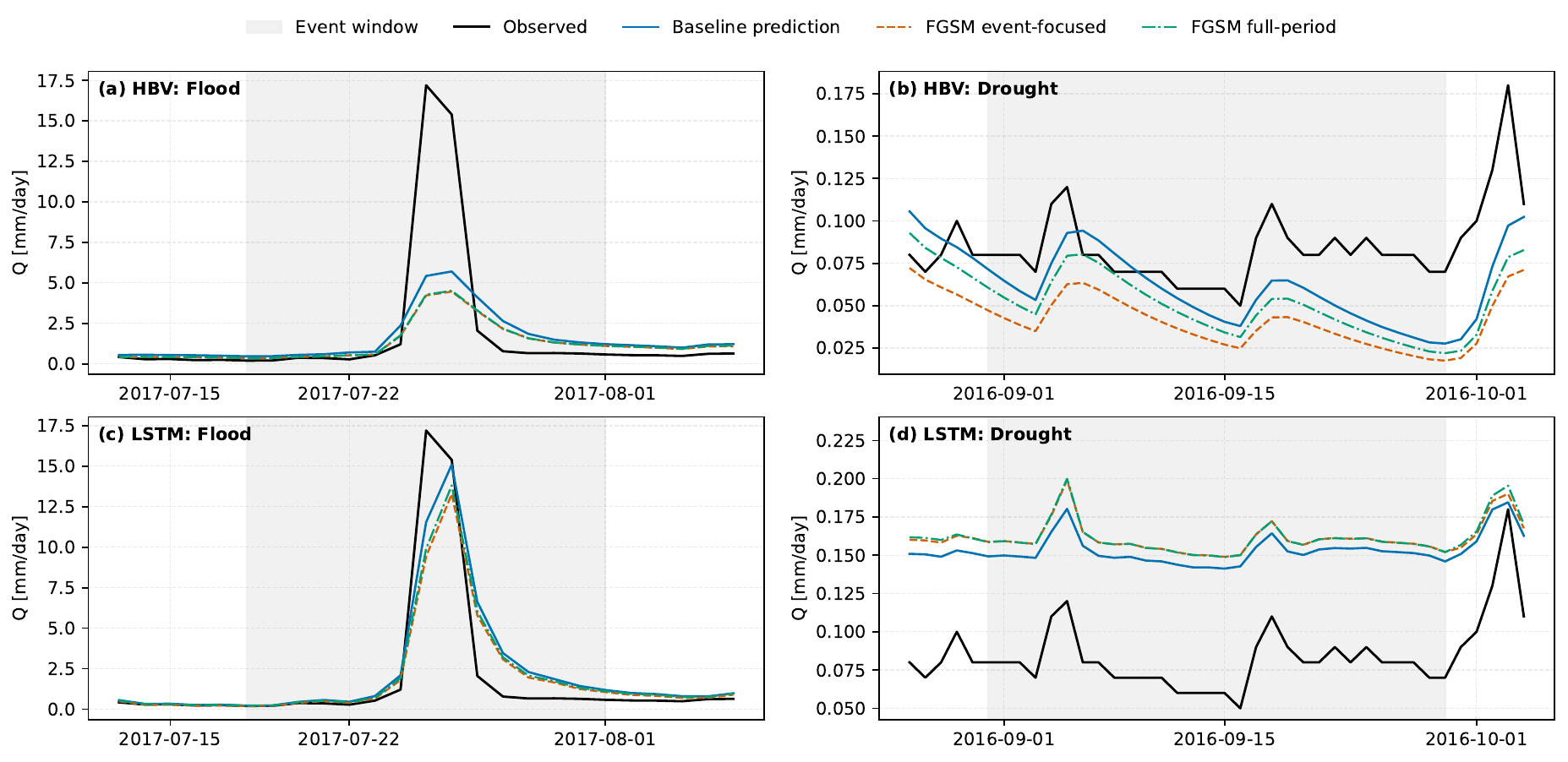}
    \caption{Event-specific robustness for CAMELS-DE catchment DE911520 under FGSM perturbations with $\epsilon=0.20$: (a) HBV during the flood event, (b) HBV during the drought event, (c) LSTM during the flood event, and (d) LSTM during the drought event.}
    \label{fig:event_robustness_DE911520}
  \end{figure}

  In addition to evaluating the effects of FGSM on overall predictive performance, we conduct an event-focused analysis for catchment DE911520 as an example. We define the flood-event window as the day of peak observed $Q$ together with the seven days before and after it, and the drought-event window as the 30-day period with the lowest mean observed $Q$. For each window, we generate FGSM perturbations using prediction error within that window as the loss and compare their effects with those of perturbations generated using the full evaluation period. As shown in Figure~\ref{fig:event_robustness_DE911520}, the event-focused and full-period perturbations produce qualitatively similar changes in the predicted hydrographs, while the event-focused perturbations generally produce greater performance degradation, with the magnitude of the difference varying among cases. This suggests that full-period FGSM perturbations can also affect predictions during the selected extreme events. However, this analysis considers only one catchment, one flood event, and one drought event; it is intended as an initial illustration rather than a comprehensive assessment of event-specific robustness.

\begin{figure}[htbp]
    \centering
    \includegraphics[width=\textwidth]{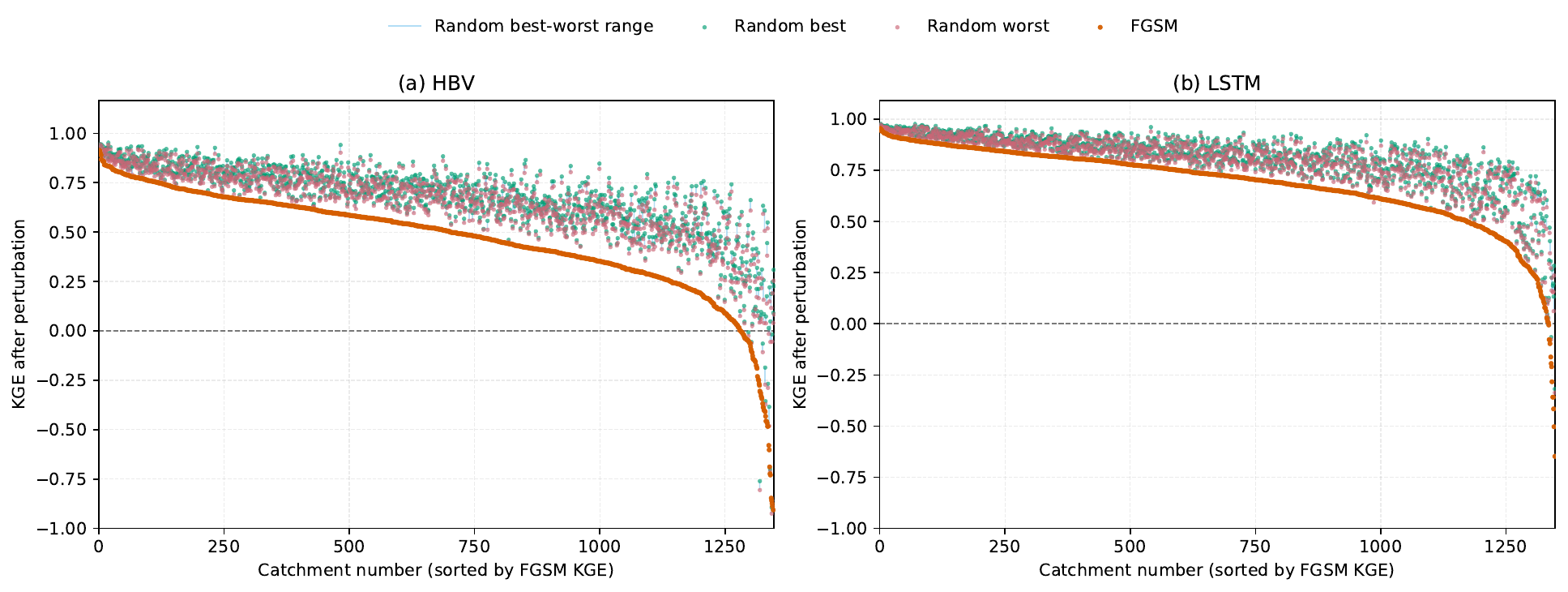}
    \caption{Comparison of KGE under FGSM and 100 random input perturbations with a component-wise perturbation budget of $\epsilon=0.2$ for (a) HBV and (b) LSTM across the CAMELS-DE catchments. Catchments are sorted by their KGE under FGSM. Orange points show the FGSM results, while green and pink points show the best and worst KGE, respectively, among the random perturbations; the light-blue lines span the corresponding best--worst ranges. The lower y-axis limit is set to $-1.0$ in both panels for visual clarity.}
    \label{fig:fgsm_vs_random_sensitivity_kge}
  \end{figure}

  To compare FGSM with a random-perturbation sensitivity analysis, we generate 100 random perturbations with the same component-wise perturbation budget ($\epsilon=0.2$). In each random trial, individual input values are randomly increased, decreased, or left unchanged, with negative $P$ and $PET$ values truncated to zero. FGSM causes substantially greater performance degradation: it produces a larger MSE increase than the worst random trial in 100\% of catchments for both models, and a larger KGE decrease in 99.7\% of LSTM catchments and 99.6\% of HBV catchments (as shown in Figure~\ref{fig:fgsm_vs_random_sensitivity_kge}). Thus, randomly sampled sensitivity perturbations can substantially understate the approximate local worst-case degradation revealed by adversarial testing.

  As a limited check, we applied the basic iterative method (BIM) of \citeA{kurakin2017adversarial} to 100 randomly selected catchments using the same MSE objective and perturbation budget as FGSM. The method produced only slightly greater performance degradation than FGSM and did not reverse the HBV--LSTM comparison within this sample. This small difference is consistent with the strong local linearity observed in the subsequent experiments. More advanced perturbation methods, such as projected gradient descent, gradient-free optimization, and physically constrained attacks, should be evaluated in future research.

  \subsection{Models' adversarial robustness at different perturbation magnitudes}

\begin{figure}[htbp]
    \centering
    \includegraphics[width=\textwidth]{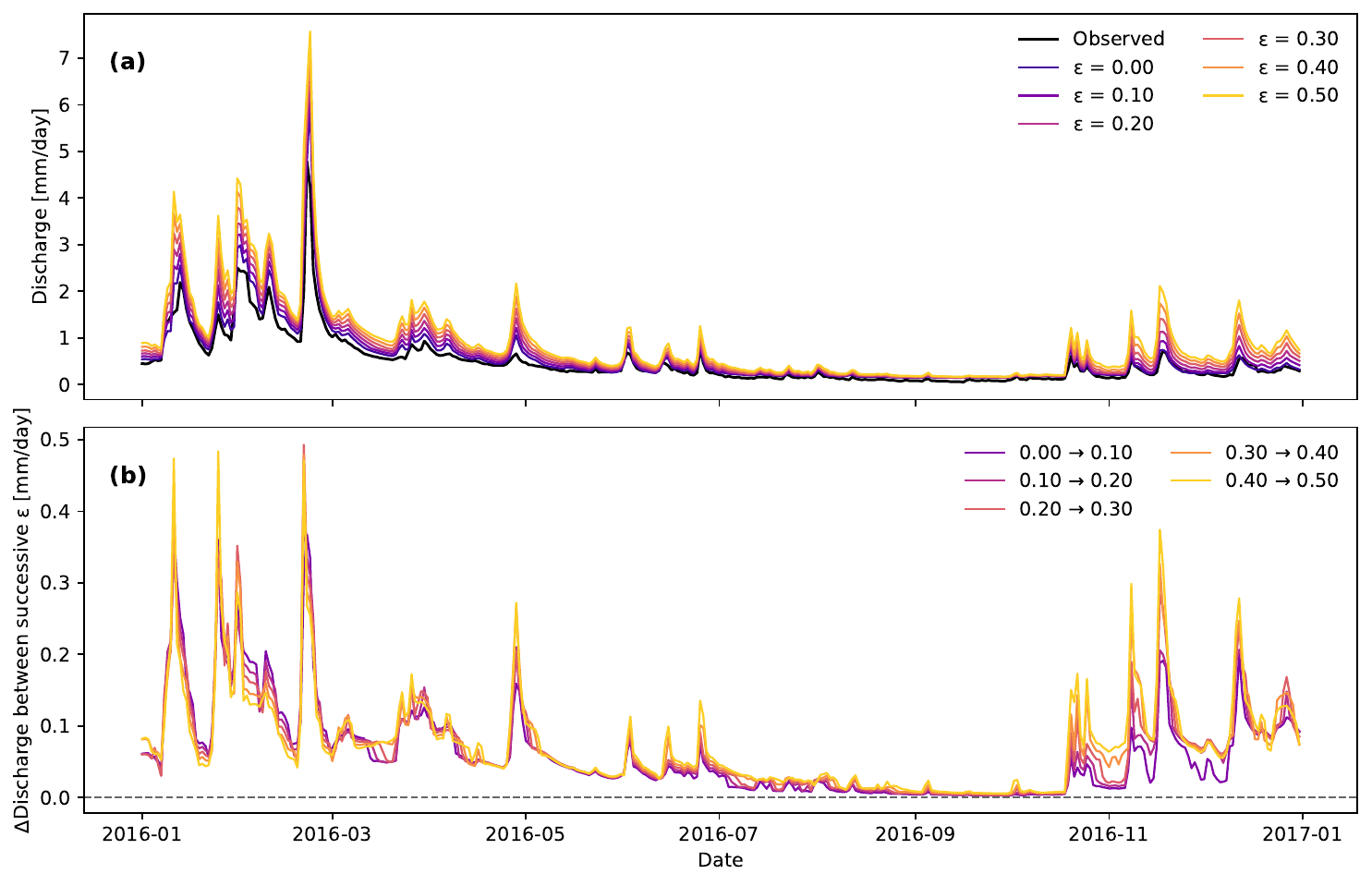}
    \caption{Effect of FGSM perturbation magnitude on LSTM discharge predictions for catchment DE911520. (a) Observed hydrograph (black) and LSTM predictions under FGSM input perturbations for $\epsilon \in \{0, 0.1, \ldots, 0.5\}$. (b) Incremental differences in predicted hydrographs between successive $\epsilon$ levels.}
    \label{figure:lstm_eps_hydrograph}
  \end{figure}

  We evaluate how model responses vary with FGSM perturbation magnitude $\epsilon$ by comparing predicted hydrographs across different $\epsilon$ values. For the LSTM model, Figure~\ref{figure:lstm_eps_hydrograph}a illustrates the predicted hydrographs for several $\epsilon$ levels in catchment DE911520, and Figure~\ref{figure:lstm_eps_hydrograph}b shows the incremental differences between successive magnitudes (for example, $\epsilon = 0.10 \rightarrow 0.20$). Overall, these incremental differences are very similar across the range of $\epsilon$ values, with a few local segments showing noticeable deviations. This indicates that the responses can be roughly approximated using linear models with respect to $\epsilon$, meaning that the perturbation effect at many time steps scales roughly proportionally with the perturbation magnitude. Some departures from perfect linearity are expected, possibly because the adversarial inputs are post-processed by clamping negative $P$ and $PET$ values to zero (a nonlinear correction step to ensure physical plausibility) and because the models themselves are nonlinear. The causes of the near-linear response will be discussed further in Section~\ref{sec:why_linearity}.

\begin{figure}[htbp]
    \centering
    \includegraphics[width=\textwidth]{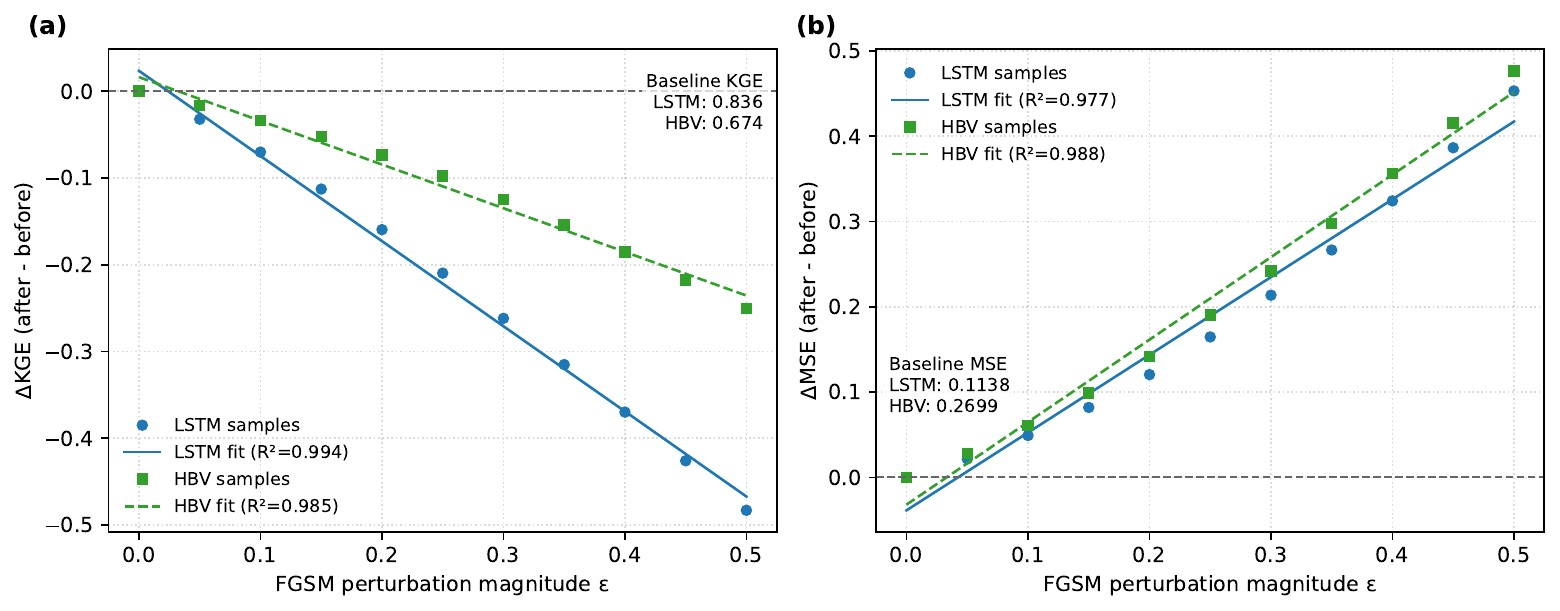}
    \caption{Relationship between FGSM perturbation magnitude $\epsilon$ and performance change for LSTM and HBV in catchment DE911520: (a) $\Delta\mathrm{KGE}$ and (b) $\Delta\mathrm{MSE}$ (after minus before). Points show model results and lines show least-squares fits with $R^2$ values. For this catchment, HBV exhibits a smaller KGE decrease but a slightly larger MSE increase than LSTM; this relative pattern may differ across catchments. Baseline performance at $\epsilon=0$ is annotated, and $\epsilon$ ranges from 0 to 0.5 in increments of 0.05.}
    \label{figure:combined_eps_metrics}
  \end{figure}

  Figure~\ref{figure:combined_eps_metrics} compares linear fits between $\epsilon$ and changes in KGE and MSE for both LSTM and HBV in catchment DE911520. Both models exhibit strong local linear relationships. For LSTM, the $R^2$ values are 0.994 for $\Delta\mathrm{KGE}$ and 0.977 for $\Delta\mathrm{MSE}$; for HBV, the corresponding values are 0.985 and 0.988. For this catchment, KGE decreases for both models, but the decrease is smaller for HBV, whereas the MSE increase is slightly larger for HBV.

  To quantify this behavior at the time-step level, we fitted a linear model $Q_t(\epsilon) \approx a_t \epsilon + b_t$ to the predicted discharge for each day $t$ in catchment DE911520 using all eleven evaluated FGSM perturbation magnitudes, $\epsilon\in\{0,0.05,0.10,\ldots,0.50\}$. We then summarized the resulting $R^2$ and slope distributions (Table~\ref{tab:fgsm_linearity_DE911520}). Both models exhibit a strong local linear response along the FGSM direction: the median $R^2$ across time steps exceeds 0.99 for both LSTM and HBV, and almost 90\% of the time steps have $R^2 \ge 0.95$. Median absolute slopes are larger for HBV than for LSTM, indicating a larger median absolute discharge-response slope for HBV in this catchment. We emphasize that this does not imply that the true input–output mapping is globally linear, and nonlinear functions (e.g., quadratic) can in fact appear nearly linear over small neighborhoods of the input space. Instead, these results show that, within the limited range of perturbations considered here, the effect of different FGSM magnitudes on the hydrograph can be well approximated and interpreted using simple linear models.

  \begin{table}[htbp]
    \centering
    \caption{Linearity summary statistics for LSTM and HBV discharge responses to FGSM perturbations for catchment DE911520. Each value summarizes fits across all time steps.}
    \label{tab:fgsm_linearity_DE911520}
    \begin{tabular}{lcccc}
      \hline
      Model & Median $R^2$ & Mean $R^2$ & $\mathrm{frac}(R^2 \ge 0.95)$ & Median $|a_t|$ \\
      \hline
      LSTM & 0.995 & 0.976 & 0.899 & 0.160 \\
      HBV  & 0.993 & 0.974 & 0.884 & 0.265 \\
      \hline
    \end{tabular}
  \end{table}

\begin{figure}[htbp]
    \centering
    \includegraphics[width=\textwidth]{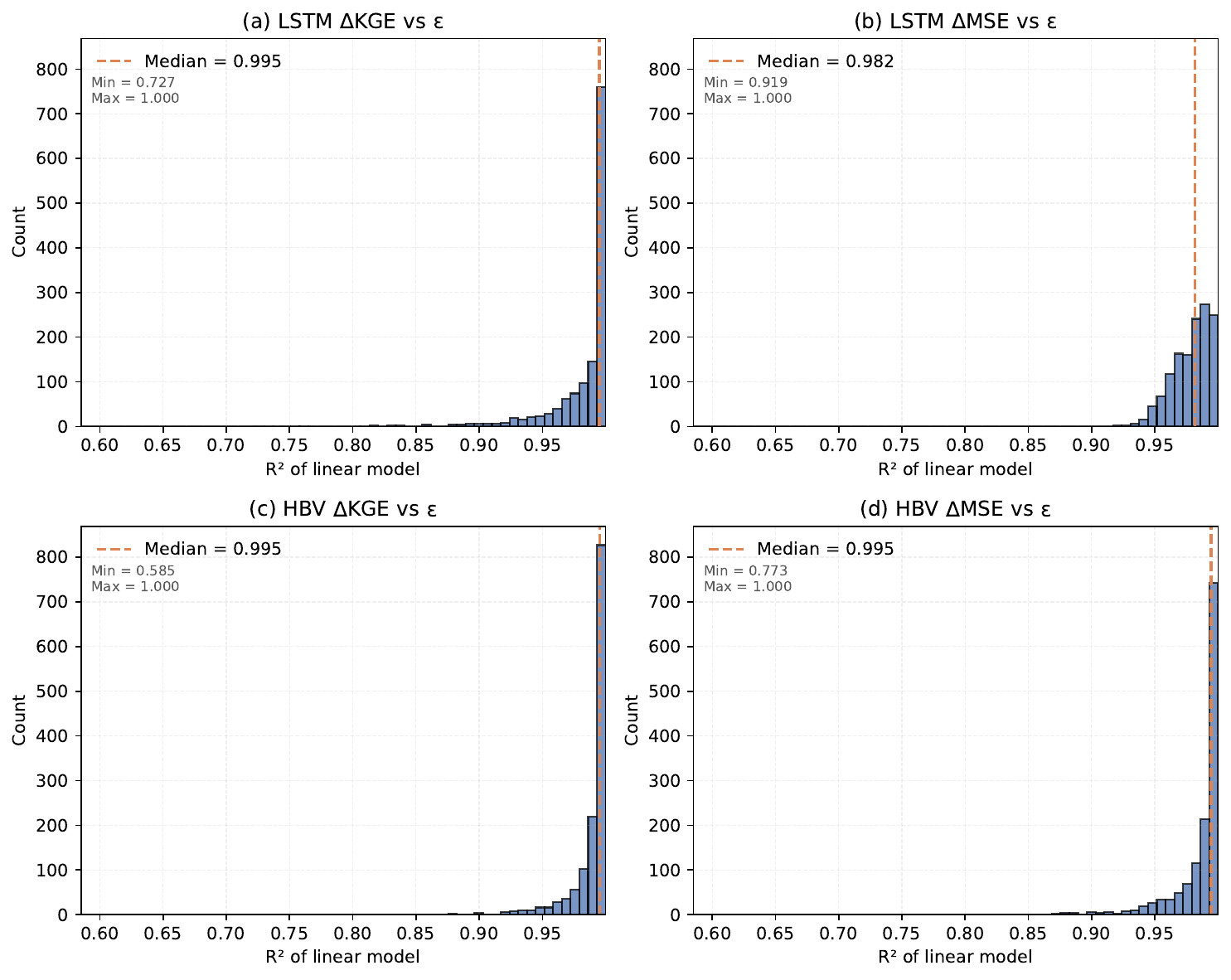}
    \caption{Histogram of $R^2$ values for linear models of performance change versus FGSM perturbation magnitude $\epsilon$ across 1,347 selected CAMELS-DE catchments. (a) LSTM $\Delta\mathrm{KGE}$ vs.\ $\epsilon$; (b) LSTM $\Delta\mathrm{MSE}$ vs.\ $\epsilon$; (c) HBV $\Delta\mathrm{KGE}$ vs.\ $\epsilon$; (d) HBV $\Delta\mathrm{MSE}$ vs.\ $\epsilon$.}
    \label{figure:r2_histogram}
  \end{figure}

\begin{figure}[htbp]
    \centering
    \includegraphics[width=\textwidth]{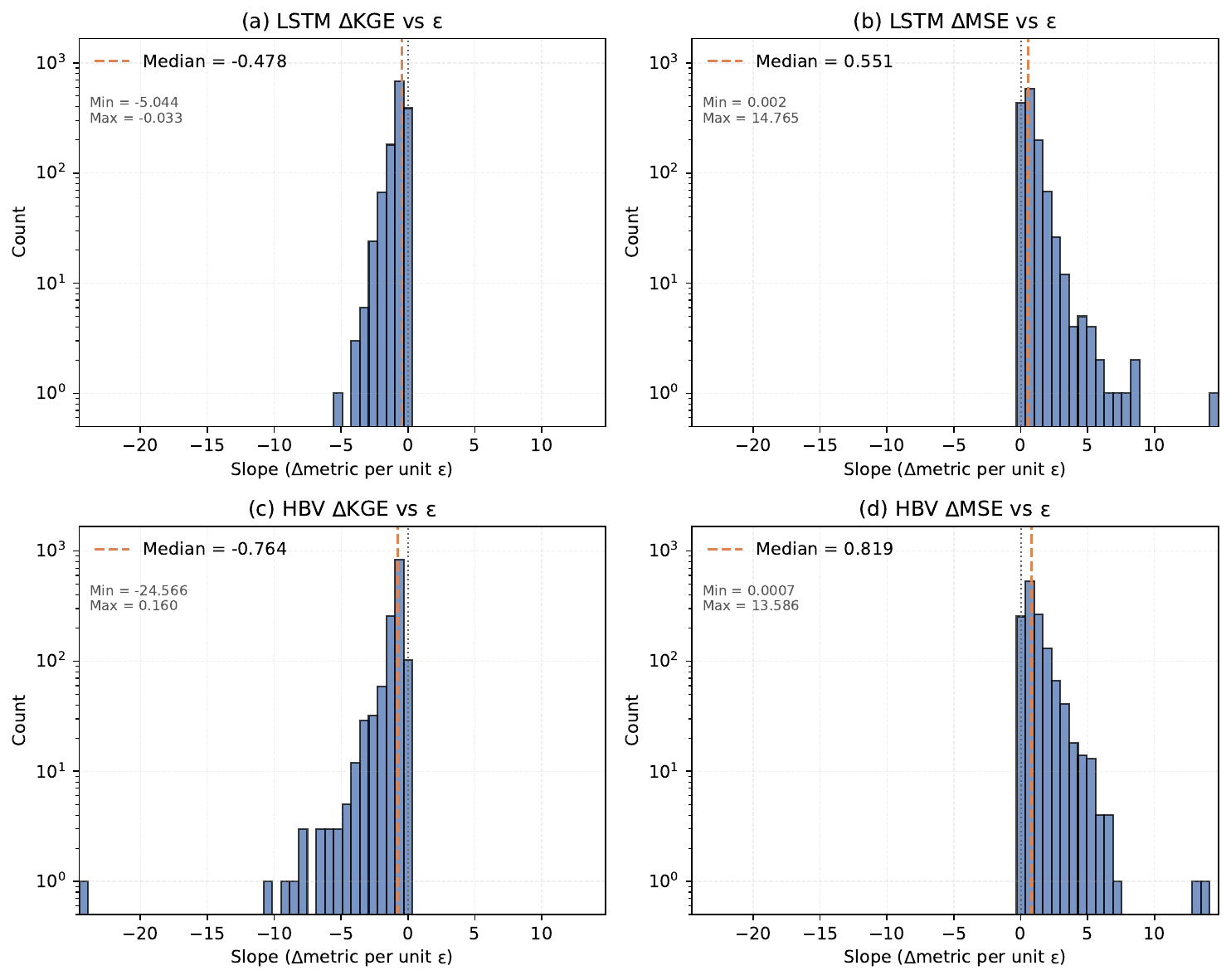}
    \caption{Histogram of slopes for linear models of performance change versus FGSM perturbation magnitude $\epsilon$ across 1,347 selected CAMELS-DE catchments. (a) LSTM $\Delta\mathrm{KGE}$ vs.\ $\epsilon$; (b) LSTM $\Delta\mathrm{MSE}$ vs.\ $\epsilon$; (c) HBV $\Delta\mathrm{KGE}$ vs.\ $\epsilon$; (d) HBV $\Delta\mathrm{MSE}$ vs.\ $\epsilon$.}
    \label{figure:slope_histogram}
  \end{figure}

  To summarize the linearity in $\epsilon$ across all catchments, we fit linear models between $\Delta\mathrm{Perf}$ and $\epsilon$ for both the LSTM and HBV models across all 1,347 CAMELS-DE catchments and report the distributions of $R^2$ and slopes in Figures~\ref{figure:r2_histogram}--\ref{figure:slope_histogram}. Across our large sample of catchments, high linearity is observed in most cases, as indicated by large $R^2$ values, with a few exceptions for both LSTM and HBV. Note that adversarial inputs in our experiments were adjusted to avoid negative values of $P$ and $PET$, and these adjustments may introduce some nonlinearity. In general, HBV exhibits larger absolute slope magnitudes than LSTM, suggesting greater sensitivity to FGSM perturbations. The median slopes for HBV are $-0.764$ for $\Delta$KGE and 0.819 for $\Delta$MSE, while the median slopes for LSTM are $-0.478$ for $\Delta$KGE and 0.551 for $\Delta$MSE. These median slopes imply that, for each 0.1 increase in perturbation magnitude, HBV exhibits approximately 0.03 more degradation than LSTM for both KGE and MSE.

  As the relationship between $\Delta\mathrm{Perf}$ and $\epsilon$ can be approximated using linear models in most evaluated cases, the slope of the fitted line provides a concise measure of a model’s sensitivity to perturbations. Under this local linear approximation, the average performance $\mathrm{RobustPerf}_{\mathrm{avg}}$ over $[\epsilon_1,\epsilon_2]$ can be approximated by the performance at the midpoint $(\epsilon_1+\epsilon_2)/2$. For nonlinear responses, it is more appropriate to compute $\mathrm{RobustPerf}_{\mathrm{avg}}$ as the normalized AUC using multiple $\epsilon$ values within the interval.

  \section{Discussion}
  \label{sec:discussion}

  \subsection{Is adversarial robustness a worthy goal?}

  Hydrological models are often used to guide high-consequence decisions: dam operations, flood warning and emergency response, water-supply scheduling, stormwater storage control, and irrigation planning. In these settings, reliability is not optional. If a model is overly sensitive to small perturbations in its inputs, then modest measurement errors, preprocessing differences, or data issues may translate into outsized changes in predicted discharge and, in turn, changes in operational decisions. Even when malicious adversaries are not a concern, adversarial robustness remains a useful goal because adversarial testing provides an approximate local worst-case assessment of model stability. \citeA{xia2025identifying} note that practitioners often distrust black-box machine learning methods despite their strong predictive performance because their predictions can be difficult to interpret. A model that can fail dramatically under imperceptible input changes will be difficult to justify in risk-averse contexts. More broadly, we would like DL rainfall–runoff models to behave more like conventional hydrologic models when confronted with small perturbations and unfamiliar conditions (e.g., nonstationarity under climate change): changes in forcing should lead to bounded, physically plausible, and interpretable changes in output rather than brittle failures. Adversarial evaluation cannot guarantee correctness under all future scenarios, but it can help reveal whether a model’s behavior is locally stable in the neighborhood of realistic forcings.

While adversarial robustness is a worthy goal, a trade-off between robustness and clean-input predictive performance can arise in some settings, such as particular adversarial-training settings \cite{raghunathan2020understanding,tsipras2019robustness}. Indeed, a model that always predicts zero streamflow would be input-invariant under these perturbations but would have poor predictive performance. Unfortunately, in regression tasks such as streamflow prediction, there is often no clear distinction between appropriate and inappropriate responses to a perturbation, though a catastrophic failure would be more noticeable. A response that satisfies basic mass-balance bounds may still be difficult to classify as appropriate or inappropriate without knowing how the real catchment would respond. Clearly, it would be desirable to have a better and more nuanced understanding of how to characterize the nature and degree of robustness of such models. While it is currently difficult to determine whether a model has the ``correct'' level of adversarial robustness, we can quantitatively compare robustness levels between methods across various tasks. We follow this line of thought in the following discussion.

  \subsection{Interpreting the adversarial robustness of common hydrological models}

  While LSTMs may be perceived as unreliable under adversarial perturbations because of DL's reputation for catastrophic failures in domains such as image recognition, our results do not indicate widespread catastrophic failure under the FGSM perturbations tested here. While HBV is not used here to define the physically correct response to a perturbation, it serves as a physically structured conceptual reference for gauging whether the LSTM exhibits disproportionately large or catastrophic predictive-performance degradation under the same FGSM framework and perturbation budget. Because LSTM degradation is generally smaller than HBV degradation, including after stratification by baseline KGE, we do not interpret the tested perturbations as revealing widespread catastrophic local failure in the LSTM. This comparison concerns predictive stability under FGSM and does not imply that the LSTM response is more physically correct or that the stronger HBV response is physically undesirable. These findings may help build confidence in LSTM-based models for risk-averse applications, but they also underline important caveats: some catchments, particularly dry catchments or those with low initial skill, can exhibit large KGE decreases under small input perturbations. At $\epsilon=0.2$, the Spearman correlations between the catchment aridity index (mean $PET$ divided by mean $P$ over 1981--2020) and the KGE decrease were 0.47 for HBV and 0.43 for LSTM, indicating a tendency toward larger KGE decreases in drier catchments. By contrast, the corresponding correlations with MSE increase were $-0.76$ and $-0.69$, respectively, underscoring the metric dependence of this relationship.

  Applications of adversarial robustness analysis in hydrology, particularly in rainfall-runoff modeling, remain relatively limited, although related efforts are emerging. \citeA{xia2025identifying} benchmarked LSTM, DeepONet, and Informer architectures for continental-scale water-quality prediction using outliers, random noise, and adversarial perturbations. Across the evaluated water-quality variables and within these architectures, variables with higher baseline performance tended to be less sensitive to adversarial perturbations. Taken together with our findings, these results suggest that robustness is not a fixed property of DL models or any single architecture; it likely depends on the prediction target, data, model, and evaluation protocol. Broader benchmarking across additional rainfall-runoff models (both conceptual and neural) and evaluation settings will be important for clarifying when and why a given model class is reliably stable.

  We ultimately do not know how a given real catchment would respond to a specified set of input perturbations. This differs from traditional adversarial testing settings, such as image classification, where a slightly perturbed image of a stop sign should still be recognized by humans and classified as a stop sign. In hydrology, simple thought experiments may provide some intuition about responses in limiting cases. For example, in an extremely dry catchment, small increases in rainfall may produce little or no runoff until a threshold is crossed, after which discharge may increase sharply, yielding a nonlinear response. By contrast, in a very wet, near-saturated catchment or a highly impervious urban catchment, additional rainfall may be converted more directly into runoff, producing an approximately linear response over a limited range. However, such thought experiments are difficult to extend beyond these extreme cases to infer how catchments under a wider range of conditions would respond to perturbations.

  For scientific understanding, hydrologists often aim to develop models whose responses to input perturbations resemble those of real-world catchments as closely as possible. However, from an operational or engineering-design perspective, there is an additional tension: we would like prediction systems to be robust to small or difficult-to-detect, potentially adversarial perturbations, so that such perturbations have minimal effects on safety, design, and operational decisions. This operational objective differs from using models to understand how natural systems respond to forcings. It underscores that there is no single ``correct'' level of robustness; the appropriate level depends on whether the goal is scientific understanding, reliable operations, or both.

  \subsection{Why does $\epsilon$-linearity emerge, and what does it tell us?}
  \label{sec:why_linearity}
  Across most catchments, changes in KGE and MSE scale approximately linearly with $\epsilon$; in the examined catchments, changes in predicted discharge are also approximately linear. We interpret this as a local approximation, not evidence of globally linear input–output mappings, and next examine the mechanisms within each model.

  \subsubsection{Linearity of LSTM models}
  In our LSTM setup, a time-distributed MLP maps the 254-dimensional LSTM hidden state at each time step to discharge. The MLP contains no dropout layers and has two hidden layers (13 and 5 units) with ReLU activations, followed by a linear output layer, i.e., $254 \rightarrow 13~(\mathrm{ReLU}) \rightarrow 5~(\mathrm{ReLU}) \rightarrow 1$ (linear). The ReLU activation function, $\mathrm{ReLU}(x) = \max(0,x)$, is piecewise linear; each ReLU unit changes between its active and inactive regimes only when its pre-activation crosses zero. Consequently, if the signs of all ReLU pre-activations in both hidden layers remain unchanged under perturbation, the MLP remains within a fixed activation region and acts as an affine mapping from the LSTM hidden state to discharge.

\begin{figure}[htbp]
    \centering
    \includegraphics[width=0.7\textwidth]{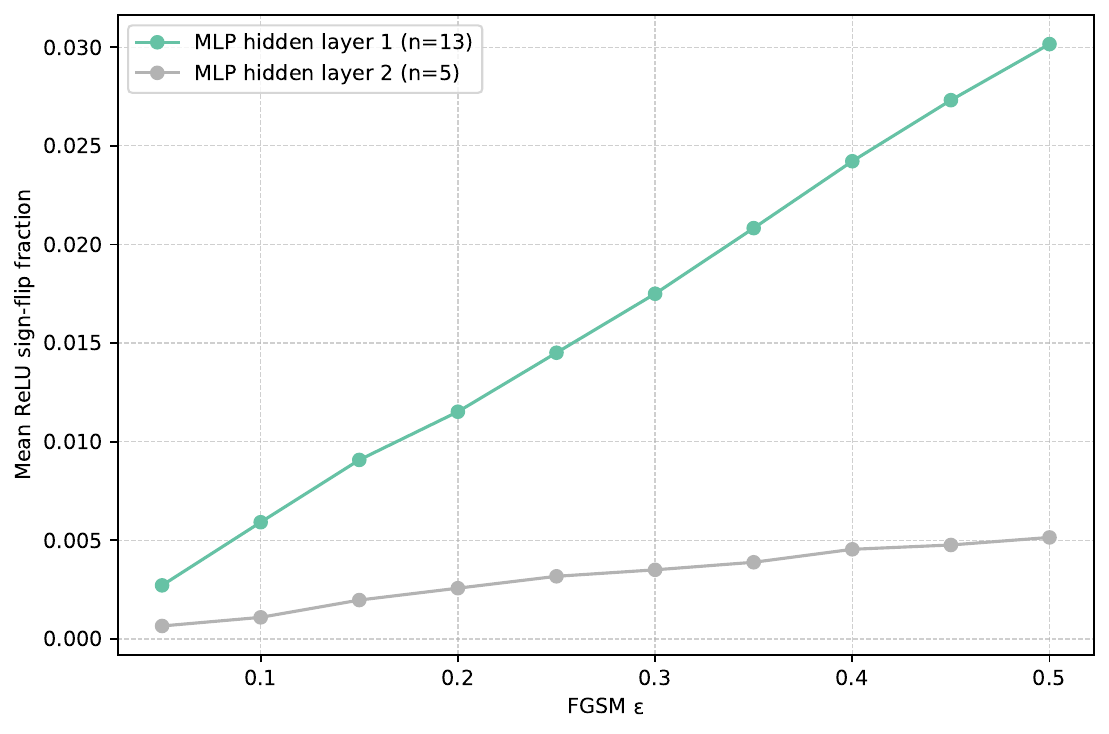}
    \caption{Mean fraction of ReLU pre-activations in each of the two MLP hidden layers whose signs differ from their unperturbed values for catchment DE911520 under different FGSM perturbation magnitudes $\epsilon$.}
    \label{figure:lstm_relu_flip}
  \end{figure}

  As a diagnostic of changes in the local activation regime, we compute, for each MLP hidden layer and test-period time step, the fraction of ReLU pre-activations whose signs differ from their unperturbed ($\epsilon=0$) values, and then average this fraction over time for catchment DE911520. Figure~\ref{figure:lstm_relu_flip} shows that the sign-flip fractions increase with $\epsilon$ but remain low overall: even at $\epsilon=0.5$, the mean fractions are approximately 3.0\% for the first hidden layer and 0.5\% for the second. These low fractions show that most individual ReLU pre-activations retain their baseline signs, indicating that the MLP mapping from the LSTM hidden state to discharge remains close to its baseline affine regime over the tested perturbation range.

\begin{figure}[htbp]
    \centering
    \includegraphics[width=\textwidth]{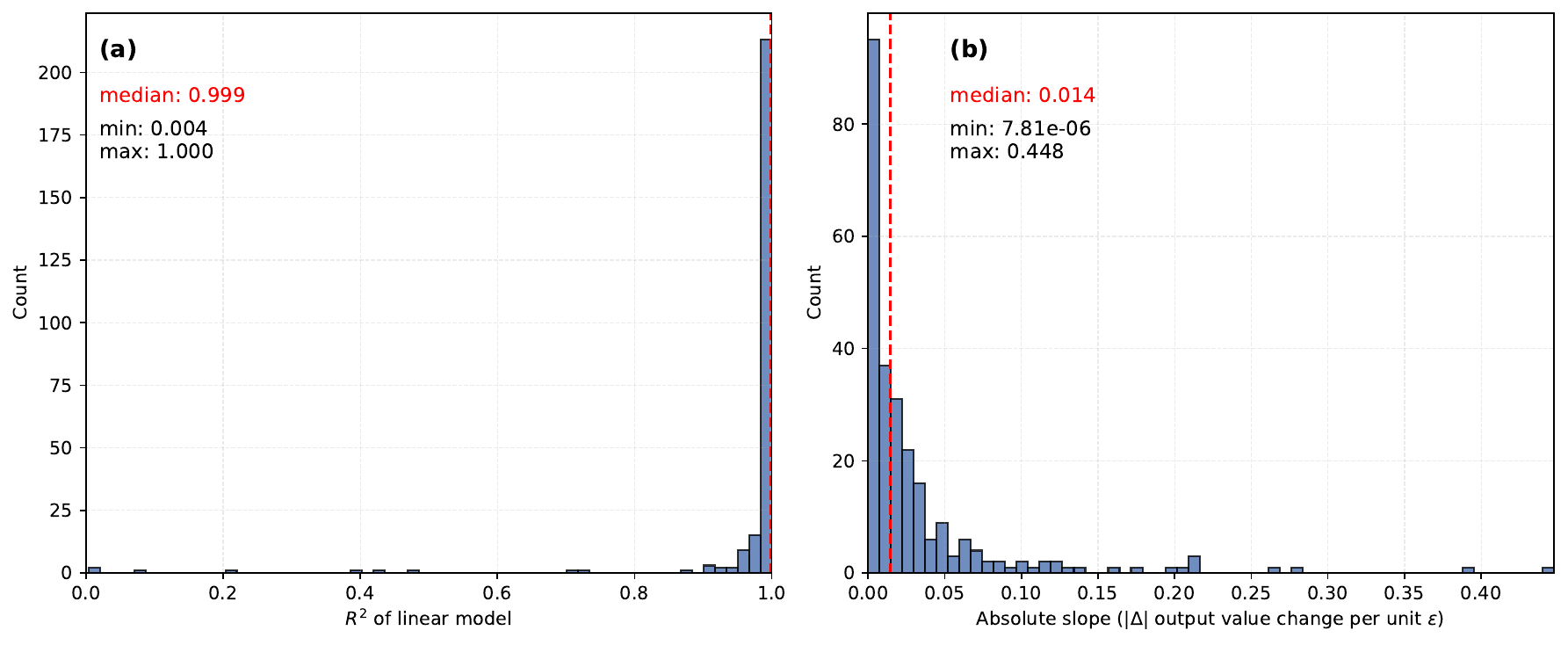}
    \caption{Histograms for catchment DE911520 summarizing per-channel linear fits of time-mean LSTM hidden-state activations versus $\epsilon$: (a) $R^2$ values and (b) absolute slopes of the linear models (change in time-mean activation per unit $\epsilon$).}
    \label{figure:activation_trends_hist}
  \end{figure}

  Having characterized the hidden-state-to-discharge mapping, we next assess whether the time-mean activations of the 254 LSTM hidden-state channels vary approximately linearly as the meteorological inputs are perturbed along the FGSM direction. For each channel, we compute its mean activation over the test period at $\epsilon$ values from 0 to 0.5 in steps of 0.05 and then fit a linear model relating mean activation to $\epsilon$. As an example, Figure~\ref{figure:activation_trends_hist}a shows the histogram of $R^2$ values from these fits for catchment DE911520. The distribution is strongly left-skewed, with most values near one and a long tail toward lower values (median $R^2 \approx 0.999$). This result suggests that, over this perturbation range, the time-mean activation of most channels is well approximated by a linear function of $\epsilon$. Figure~\ref{figure:activation_trends_hist}b shows the corresponding distribution of absolute slopes, which is strongly right-skewed, with most values near zero and a long tail toward larger slopes.

  Together, these distributions indicate that, along the FGSM direction, most time-mean hidden-state activations vary approximately linearly with $\epsilon$, while a relatively small subset of channels is more sensitive. We examined several catchments and found qualitatively similar patterns, although the detailed responses varied across catchments. The present analysis reports only the time-mean activations of the LSTM hidden-state channels; their responses at individual time steps should be examined in future work. This local result does not imply global linearity; for example, \citeA{baste2025unveiling} showed that the examined LSTMs saturated at empirical design limits below their theoretical upper limits under extreme forcing.

  \subsubsection{Linearity of HBV models}
\begin{figure}[htbp]
    \centering
    \includegraphics[width=\textwidth]{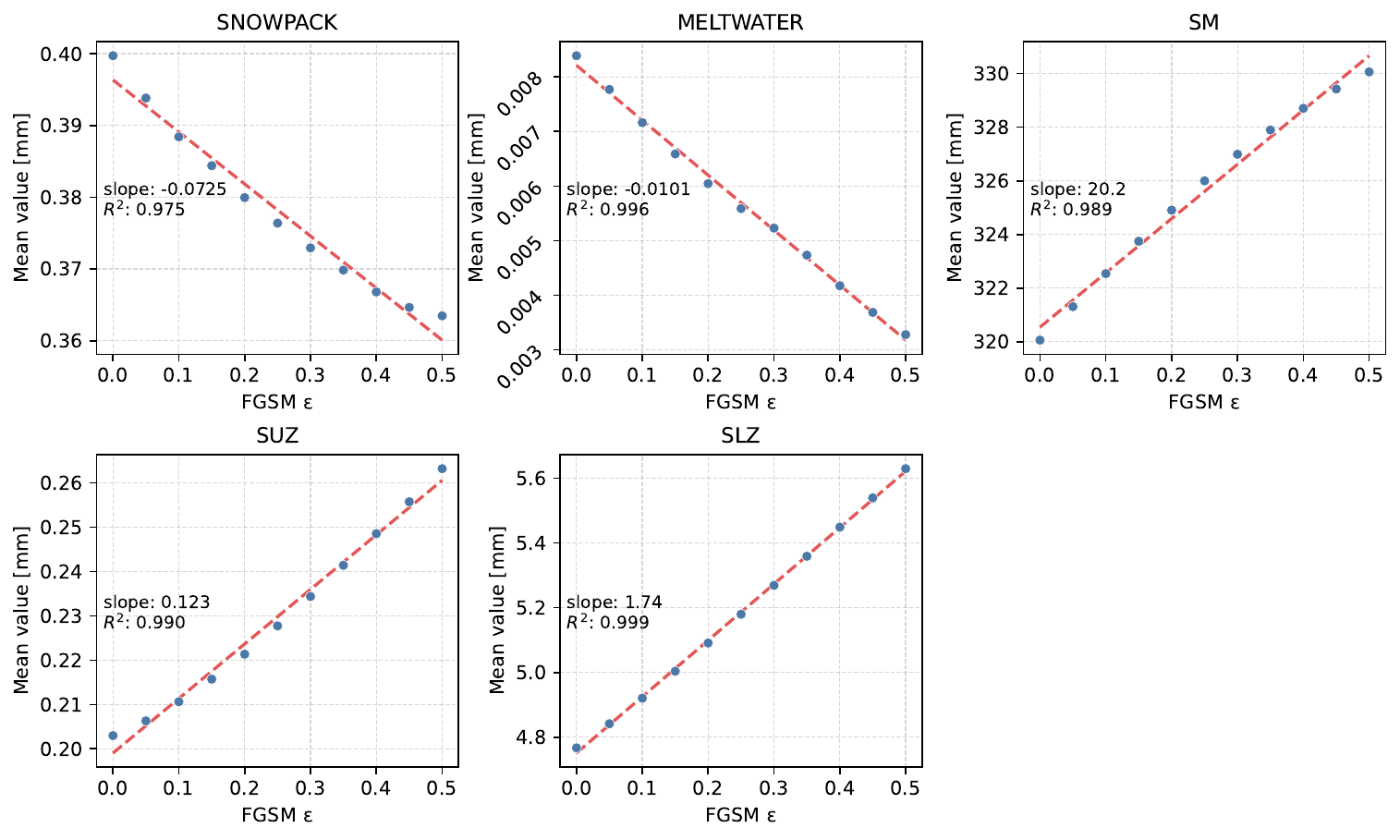}
    \caption{Mean storage of the five HBV stores for catchment DE911520 during the test period (excluding warm-up) at different FGSM perturbation magnitudes $\epsilon$. Each panel shows the mean-storage values, the fitted linear model (dashed line), and the slope and $R^2$ of the fit.}
    \label{figure:hbv_internal_response_grid_DE911520}
  \end{figure}

  To examine HBV's internal response to FGSM, we compute, for each store, its mean value across the test period (excluding warm-up) for a range of $\epsilon$ values and fit a linear model relating mean storage to $\epsilon$. Figure~\ref{figure:hbv_internal_response_grid_DE911520} shows the results for catchment DE911520. Across the five stores, mean storage is generally well approximated by a linear function of $\epsilon$ (high $R^2$ values), despite differences in storage magnitude, ranging from submillimeter values for meltwater (MW) to hundreds of millimeters for soil moisture (SM). For this catchment, the largest fitted slope is 20.2~mm per unit $\epsilon$ for SM, corresponding to an approximately 10~mm change in mean SM over the tested range from $\epsilon=0$ to 0.5. However, the relationship between mean storage and $\epsilon$ can be nonlinear for certain stores and catchments. Moreover, because the analysis uses time-averaged storage, a high $R^2$ does not imply that the daily storage trajectory is linear in $\epsilon$.

  Why do such near-linear relationships appear for some catchments? Several HBV state-update and flux calculations are linear or piecewise linear, whereas other components are smooth nonlinear functions (e.g., $(\cdot)^{\mathrm{BETA}}$). Under small, fixed-direction perturbations (FGSM along a single gradient-sign direction), the model may remain within the same process regimes for most time steps. Downstream of these state and flux calculations, unit-hydrograph routing is a fixed linear operator, so any approximately linear dependence of generated runoff on $\epsilon$ is preserved in the routed discharge. Thus, the combination of linear or piecewise-linear process components, locally smooth nonlinear components away from regime thresholds, and linear routing helps explain why simple linear models provide a good local approximation to the HBV response to small FGSM perturbations.

  These results pertain to FGSM perturbations within a local neighborhood of the baseline meteorological forcings. Model behavior under larger perturbations or out-of-distribution inputs warrants further study.

  \subsubsection{Beyond the FGSM direction: $\epsilon$-linearity along random directions?}
  If the discharge response is locally linear along the FGSM direction, the response at output time step $t$ can be expressed as
  \[
    f_t\left(\operatorname{adjust}\!\left(\mathbf{X} + \epsilon \mathbf{D}_{\mathrm{FGSM}}\right)\right)
    \approx b_t+\epsilon k_t,
  \]
  where $f_t(\cdot)$ denotes the specified model's predicted discharge at output time step $t$ for a given meteorological input sequence; $\mathbf{X}$ denotes the baseline meteorological input sequence; $\mathbf{D}_{\mathrm{FGSM}}\in\{-1,0,1\}^{F\times m}$ is the element-wise sign of the loss gradient for the specified model, input sequence, and target discharge, with $F$ input variables and $m$ input time steps; and $k_t$ and $b_t$ are the fitted slope and intercept, respectively, at output time step $t$. Because the fit includes $\epsilon=0$, $b_t$ is expected to be close to the unperturbed prediction $f_t(\mathbf{X})$. Here, $\operatorname{adjust}(\cdot)$ sets any negative perturbed $P$ and $PET$ values to zero while leaving the perturbed $T$ values unchanged.

  To examine directions other than the FGSM direction, we replace $\mathbf{D}_{\mathrm{FGSM}}$ with random sign matrices in $\{-1,0,1\}^{F\times m}$. For the illustrative catchment shown below, we sample 100 such directions, vary $\epsilon$ over the tested range, and, at each output time step of the test period, fit a linear model relating predicted discharge to $\epsilon$ for both LSTM and HBV. For each random direction, we summarize the linearity using the median $R^2$, the median absolute slope, and the fraction of time steps with $R^2\geq0.95$. We also repeat this analysis for other catchments as a consistency check.

  \begin{figure}[htbp]
    \centering
    \includegraphics[width=\textwidth]{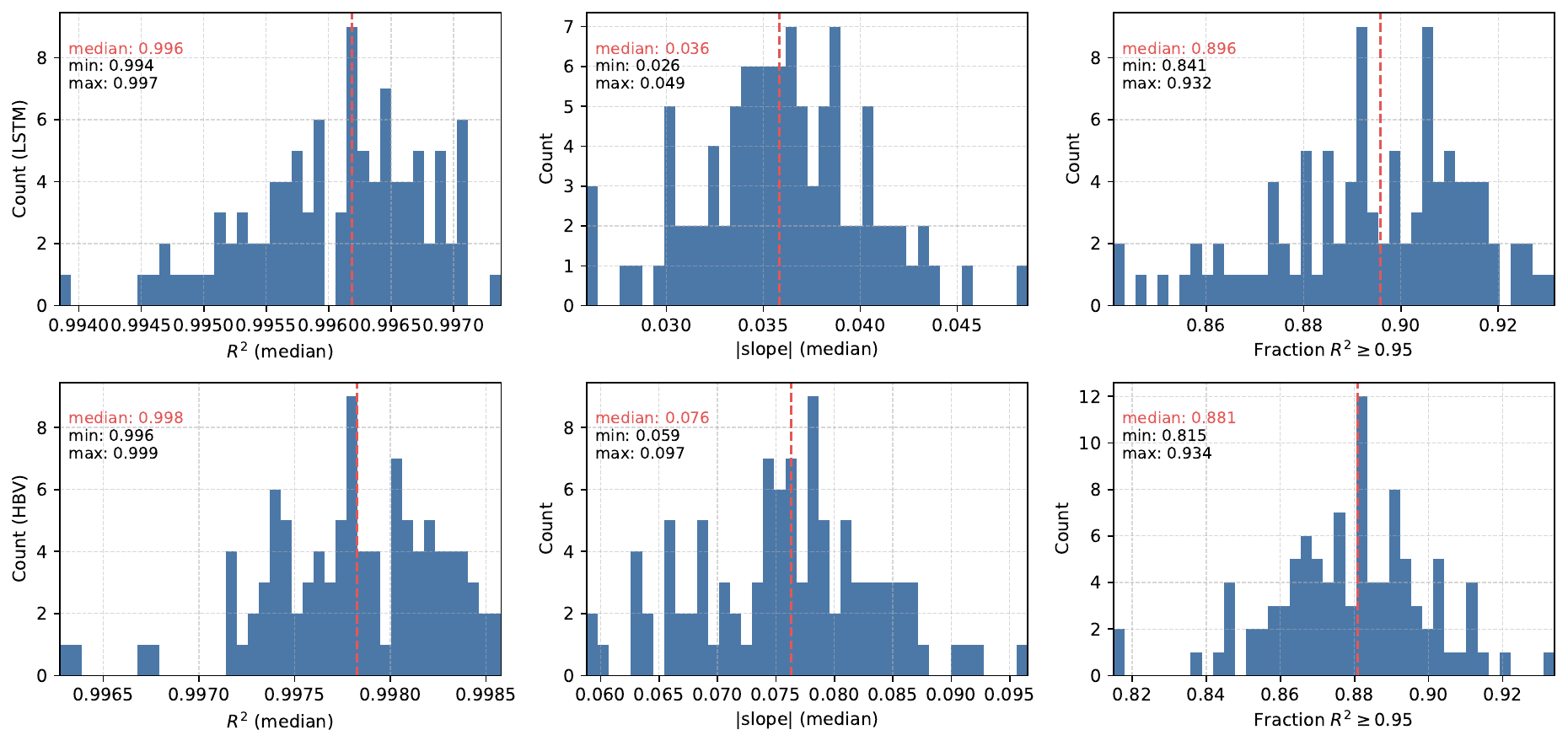}
    \caption{Random-direction linearity for the illustrative catchment DE911520. For each random direction, the histograms show (left) the median $R^2$ of the discharge–$\epsilon$ fits, (middle) the median absolute slope, and (right) the fraction of time steps with $R^2\geq 0.95$ for LSTM (top row) and HBV (bottom row).}
    \label{figure:random_dir_linearity_hist_DE911520}
  \end{figure}

  Figure \ref{figure:random_dir_linearity_hist_DE911520} shows the results for the illustrative catchment DE911520. For this catchment, the median $R^2$ exceeds 0.99 for every sampled direction, and the minimum fraction of time steps with $R^2\geq0.95$ across all sampled directions and both models is approximately 81.5\%. Within each model, the median absolute slopes vary over a relatively narrow range across directions and are generally larger for HBV than for LSTM. The other tested catchments show qualitatively consistent patterns. Thus, for DE911520 and the other tested catchments over the tested perturbation range, $\epsilon$-linearity is not restricted to the FGSM direction. The fitted slope represents a direction- and state-dependent local sensitivity determined jointly by the model parameters, baseline inputs and internal states, and perturbation direction.

  \subsection{Future work}

  The results presented in this study were obtained using a single-step, gradient-based adversarial perturbation method that probes model characteristics in the vicinity of realistic forcings. Future studies should evaluate other testing scenarios, including iterative and gradient-free perturbation methods, larger perturbations that cross model thresholds, and nonstationary or out-of-distribution inputs. These tests would help establish where local linearity breaks down and how model vulnerability changes across input regimes. Cross-forcing evaluations using independent meteorological products such as Caravan-DE, as well as operational forecast forcings with potentially larger and structured biases, are needed before the present conclusions can be extended to operational prediction settings.

  We repeated the analysis using a separately trained regional LSTM in which a standardized 15-dimensional vector of climatic, topographic, land-cover, and soil attributes was repeated at each time step and concatenated with $P$, $T$, and $PET$, replacing the learned 16-dimensional catchment-ID embedding. This variant produced similar qualitative results: at $\epsilon=0.20$, it exhibited more performance degradation than the learned-embedding LSTM but less than HBV, and its performance degraded approximately linearly with $\epsilon$ over the tested range. These results suggest that the main findings are not unique to the catchment-embedding formulation of LSTM, although a broader comparison of model architectures is still needed. Future studies should extend the evaluation to other learning-based hydrological architectures, including temporal convolutional networks (TCNs) \cite{xu2021tcn}, differentiable HBV (dHBV) frameworks \cite{feng2022differentiable,AcunaEspinoza2024}, and Transformer architectures \cite{liu2025rnns}.

  Future controlled comparisons should fit both model classes using matched data periods and objectives and should assess local and regional fitting strategies to better isolate the contribution of model architecture to robustness.

  Although we compared the robustness of HBV and LSTM models, we do not know how a specific real-world catchment would respond to analogous input perturbations. We therefore cannot determine whether the observed levels of model robustness are a ``feature'' or a ``bug.'' Future work should examine how perturbation-induced changes in discharge predictions affect flood warnings, water-supply decisions, and risk assessments, and should relate model responses to real-world catchment behavior.

  The FGSM experiments in this study are theoretical white-box stress tests and are not intended to simulate realistic operational cyberattacks. They assume access to model gradients and observed discharge labels, whereas operational adversaries may have only black-box access and face other practical constraints. Future studies should evaluate these more realistic settings and examine threats to environmental sensor readings and data transmissions, including data manipulation or loss, as well as Trojan or backdoor attacks on real-time systems. Strategies for improving model robustness should also be investigated. Forcing fusion with adaptively weighted input sources has improved predictive performance \cite{sawadekar2025improving}, but whether it improves resilience to corrupted or missing inputs remains to be tested.

  \section{Conclusions}
  \label{sec:conclusion}

  This study evaluated the performance of a regional LSTM model and an HBV model across 1,347 CAMELS-DE catchments under FGSM adversarial perturbations of different magnitudes $\epsilon$. The main conclusions are:

  \begin{enumerate}
    \item Across most catchments, FGSM perturbations reduce KGE and increase MSE. While we did not observe catastrophic failure overall, the performance degradation is systematic enough to merit attention in robustness assessment and reporting.
    \item For the configurations examined, HBV generally exhibits greater predictive-performance degradation than LSTM under FGSM perturbations of different magnitudes. Even after perturbation, LSTM can retain better predictive performance than HBV. However, the models were fit using different periods, objectives, and local versus regional strategies, so the observed differences cannot be attributed solely to model architecture.

    \item In the majority of catchments, changes in KGE and MSE for both LSTM and HBV are approximately linear in perturbation magnitude $\epsilon$ over the tested range. For some catchments, predicted discharge at many individual time steps also scales approximately linearly with $\epsilon$. These findings provide useful local approximations of the perturbation effects but do not imply that the underlying input–output mappings are globally linear.
    \item For some catchments, randomly sampled perturbation directions were also tested over a range of $\epsilon$ values, with individual input values increased, decreased, or left unchanged. For both LSTM and HBV, discharge remained well approximated by a linear function of $\epsilon$ along most sampled directions, indicating that, in the vicinity of realistic forcings, local $\epsilon$-linearity was not limited to the FGSM direction.
  \end{enumerate}

  Overall, HBV and LSTM exhibit qualitatively similar response patterns under the tested perturbations. Using HBV as a physically structured conceptual reference (rather than as ground truth), we find no evidence that LSTM exhibits disproportionately large or widespread catastrophic failure under these local FGSM stress tests. Whether such behavior should be regarded as a ``feature'' that aids interpretation or a ``bug'' that exposes vulnerabilities depends on the application context and the physical characteristics of the catchment under study. Theory-informed expectations of catchment behavior under analogous perturbations are needed to judge whether the modeled responses are desirable. Nonetheless, our results provide a starting point for systematic robustness evaluation and clarify how different types of hydrological models respond locally to controlled input perturbations.

  \acknowledgments
  We thank Omar Wani and Majid Bayati for helpful discussions. Yang Yang was supported by the Seed Fund for Basic Research for New Staff from the University Research Committee of The University of Hong Kong (HKU) when he worked at HKU. Joseph Janssen was supported by an NSERC PhD fellowship as well as the Thompson Postdoctoral Fellowship in Geophysics of Stanford University. AI tools were used to assist with polishing the writing of the manuscript. AI coding tools were used to assist with code development, including reviewing, cleaning up, organizing, and improving the original source code written by the authors, and were also used to assist with developing some numerical-modeling workflows and analyzing their results. The authors reviewed all AI-assisted changes and analyses, and take full responsibility for the final manuscript and code. We thank the Editor, Associate Editor, and anonymous reviewers of Water Resources Research for their constructive comments, which helped improve the manuscript. This document was created using a LaTeX template from the American Geophysical Union (AGU) journals. The data used in this study are available as the CAMELS-DE dataset (Version 1.0.0; DOI: 10.5281/zenodo.13837553) \cite{camelsde_zenodo_13837553,loritz2024camels}. The source code used to run the experiments and generate the figures is available at \url{https://github.com/stsfk/hydrological_adversarial_robustness}.

  \bibliography{bibliography}

@article{andreassian2009hess,
  title     = {HESS Opinions" Crash tests for a standardized evaluation of hydrological models"},
  author    = {Andr{\'e}assian, Vazken and Perrin, Charles and Berthet, Lionel and Le Moine, Nicolas and Lerat, Julien and Loumagne, C{\'e}cile and Oudin, Ludovic and Mathevet, Thibault and Ramos, M-H and Val{\'e}ry, Audrey},
  journal   = {Hydrology and Earth System Sciences},
  volume    = {13},
  number    = {10},
  pages     = {1757--1764},
  year      = {2009},
  publisher = {Copernicus GmbH}
}

@misc{Yang2024generative_lumped,
  title         = {Learning Generative Models for Lumped Rainfall-Runoff Modeling},
  author        = {Yang Yang and Ting Fong May Chui},
  year          = {2024},
  eprint        = {2309.09904},
  archiveprefix = {arXiv},
  primaryclass  = {physics.geo-ph},
  url           = {https://arxiv.org/abs/2309.09904}
}

@inproceedings{akiba2019optuna,
  title     = {Optuna: A next-generation hyperparameter optimization framework},
  author    = {Akiba, Takuya and Sano, Shotaro and Yanase, Toshihiko and Ohta, Takeru and Koyama, Masanori},
  booktitle = {Proceedings of the 25th ACM SIGKDD international conference on knowledge discovery \& data mining},
  pages     = {2623--2631},
  year      = {2019}
}

@incollection{shen2021applications,
  author    = {Shen, Chaopeng and Lawson, Kathryn},
  title     = {Applications of Deep Learning in Hydrology},
  booktitle = {Deep Learning for the Earth Sciences: A Comprehensive Approach to Remote Sensing, Climate Science, and Geosciences},
  editor    = {Camps-Valls, Gustau and Tuia, Devis and Zhu, Xiao Xiang and Reichstein, Markus},
  chapter   = {19},
  pages     = {283--297},
  publisher = {Wiley},
  year      = {2021},
  doi       = {10.1002/9781119646181.ch19},
  url       = {https://doi.org/10.1002/9781119646181.ch19}
}

@article{liu2025rnns,
  author  = {Liu, J. and Shen, C. and O'Donncha, F. and Song, Y. and Zhi, W. and Beck, H. E. and Bindas, T. and Kraabel, N. and Lawson, K.},
  title   = {From RNNs to Transformers: benchmarking deep learning architectures for
             hydrologic prediction},
  journal = {Hydrology and Earth System Sciences},
  volume  = {29},
  year    = {2025},
  number  = {23},
  pages   = {6811--6828},
  url     = {https://hess.copernicus.org/articles/29/6811/2025/},
  doi     = {10.5194/hess-29-6811-2025}
}

@article{xu2021tcn,
  author   = {Xu, Yuanhao and Hu, Caihong and Wu, Qiang and Li, Zhichao and Jian, Shengqi and Chen, Youqian},
  title    = {Application of temporal convolutional network for flood forecasting},
  journal  = {Hydrology Research},
  volume   = {52},
  number   = {6},
  pages    = {1455-1468},
  year     = {2021},
  month    = {07},
  issn     = {0029-1277},
  doi      = {10.2166/nh.2021.021},
  url      = {https://doi.org/10.2166/nh.2021.021},
  eprint   = {https://iwaponline.com/hr/article-pdf/52/6/1455/1517522/nh0521455.pdf}
}

@misc{KIT-Hy2DL_github,
  author       = {Acu\~na Espinoza, Eduardo and Loritz, Ralf and {\'A}lvarez Chaves, Manuel},
  title        = {KIT-HYD/Hy2DL},
  howpublished = {GitHub repository, branch main, commit 8bc28e7},
  year         = {2025},
  note         = {Accessed: 2025-12-10},
  url          = {https://github.com/KIT-HYD/Hy2DL/tree/main}
}

@article{AcunaEspinoza2024,
  author  = {Acu\~na Espinoza, E. and Loritz, R. and \'Alvarez Chaves, M. and B\"auerle, N. and Ehret, U.},
  title   = {To bucket or not to bucket? Analyzing the performance and interpretability of hybrid hydrological models with dynamic parameterization},
  journal = {Hydrology and Earth System Sciences},
  volume  = {28},
  year    = {2024},
  number  = {12},
  pages   = {2705--2719},
  url     = {https://hess.copernicus.org/articles/28/2705/2024/},
  doi     = {10.5194/hess-28-2705-2024}
}

@article{carlini2019evaluating,
  title   = {On evaluating adversarial robustness},
  author  = {Carlini, Nicholas and Athalye, Anish and Papernot, Nicolas and Brendel, Wieland and Rauber, Jonas and Tsipras, Dimitris and Goodfellow, Ian and Madry, Aleksander and Kurakin, Alexey},
  journal = {arXiv preprint arXiv:1902.06705},
  year    = {2019}
}

@article{tyralis2019brief,
  title     = {A brief review of random forests for water scientists and practitioners and their recent history in water resources},
  author    = {Tyralis, Hristos and Papacharalampous, Georgia and Langousis, Andreas},
  journal   = {Water},
  volume    = {11},
  number    = {5},
  pages     = {910},
  year      = {2019},
  publisher = {MDPI}
}

@article{nearing2021role,
  title     = {What role does hydrological science play in the age of machine learning?},
  author    = {Nearing, Grey S and Kratzert, Frederik and Sampson, Alden Keefe and Pelissier, Craig S and Klotz, Daniel and Frame, Jonathan M and Prieto, Cristina and Gupta, Hoshin V},
  journal   = {Water Resources Research},
  volume    = {57},
  number    = {3},
  pages     = {e2020WR028091},
  year      = {2021},
  publisher = {Wiley Online Library}
}

@article{xu2021machine,
  title     = {Machine learning for hydrologic sciences: An introductory overview},
  author    = {Xu, Tianfang and Liang, Feng},
  journal   = {Wiley Interdisciplinary Reviews: Water},
  volume    = {8},
  number    = {5},
  pages     = {e1533},
  year      = {2021},
  publisher = {Wiley Online Library}
}

@article{sawadekar2025improving,
  title     = {Improving differentiable hydrologic modeling with interpretable forcing fusion},
  author    = {Sawadekar, Kamlesh and Song, Yalan and Pan, Ming and Beck, Hylke and McCrary, Rachel and Ullrich, Paul and Lawson, Kathryn and Shen, Chaopeng},
  journal   = {Journal of Hydrology},
  volume    = {659},
  pages     = {133320},
  year      = {2025},
  publisher = {Elsevier}
}

@article{yang2021reliability,
  title     = {Reliability assessment of machine learning models in hydrological predictions through metamorphic testing},
  author    = {Yang, Yang and Chui, Ting Fong May},
  journal   = {Water Resources Research},
  volume    = {57},
  number    = {9},
  pages     = {e2020WR029471},
  year      = {2021},
  publisher = {Wiley Online Library}
}

@article{reichert2024metamorphic,
  title     = {Metamorphic testing of machine learning and conceptual hydrologic models},
  author    = {Reichert, Peter and Ma, Kai and H{\"o}ge, Marvin and Fenicia, Fabrizio and Baity-Jesi, Marco and Feng, Dapeng and Shen, Chaopeng},
  journal   = {Hydrology and Earth System Sciences},
  volume    = {28},
  number    = {11},
  pages     = {2505--2529},
  year      = {2024},
  publisher = {Copernicus GmbH}
}

@article{kratzert2018rainfall,
  title     = {Rainfall--runoff modelling using long short-term memory (LSTM) networks},
  author    = {Kratzert, Frederik and Klotz, Daniel and Brenner, Claire and Schulz, Karsten and Herrnegger, Mathew},
  journal   = {Hydrology and Earth System Sciences},
  volume    = {22},
  number    = {11},
  pages     = {6005--6022},
  year      = {2018},
  publisher = {Copernicus Publications G{\"o}ttingen, Germany}
}

@article{kratzert2019towards,
  author    = {Kratzert, Frederik and Klotz, Daniel and Shalev, Guy and Klambauer, G{\"u}nter and Hochreiter, Sepp and Nearing, Grey},
  title     = {Towards Learning Universal, Regional, and Local Hydrological Behaviors via Machine Learning Applied to Large-Sample Datasets},
  journal   = {Hydrology and Earth System Sciences},
  volume    = {23},
  number    = {12},
  pages     = {5089--5110},
  year      = {2019},
  doi       = {10.5194/hess-23-5089-2019},
  url       = {https://doi.org/10.5194/hess-23-5089-2019},
  publisher = {Copernicus Publications}
}

@article{tuptuk2021systematic,
  title     = {A systematic review of the state of cyber-security in water systems},
  author    = {Tuptuk, Nilufer and Hazell, Peter and Watson, Jeremy and Hailes, Stephen},
  journal   = {Water},
  volume    = {13},
  number    = {1},
  pages     = {81},
  year      = {2021},
  publisher = {MDPI}
}

@inproceedings{tsipras2019robustness,
  title     = {Robustness May Be at Odds with Accuracy},
  author    = {Tsipras, Dimitris and Santurkar, Shibani and Engstrom, Logan and Turner, Alexander and Madry, Aleksander},
  booktitle = {International Conference on Learning Representations},
  number    = {2019},
  year      = {2019}
}

@article{xia2025identifying,
  author    = {Xia, Xiaobo
               and Liu, Xiaofeng
               and Liu, Jiale
               and Fang, Kuai
               and Lu, Lu
               and Oymak, Samet
               and Currie, William S.
               and Liu, Tongliang},
  title     = {Identifying trustworthiness challenges in deep learning models for continental-scale water quality prediction},
  journal   = {Nexus},
  year      = {2025},
  month     = {Dec},
  day       = {16},
  publisher = {Elsevier},
  volume    = {2},
  number    = {4},
  issn      = {2950-1601},
  doi       = {10.1016/j.ynexs.2025.100104},
  url       = {https://doi.org/10.1016/j.ynexs.2025.100104}
}

@inproceedings{raghunathan2020understanding,
  title        = {Understanding and Mitigating the Tradeoff between Robustness and Accuracy},
  author       = {Raghunathan, Aditi and Xie, Sang Michael and Yang, Fanny and Duchi, John and Liang, Percy},
  booktitle    = {International Conference on Machine Learning},
  pages        = {7909--7919},
  year         = {2020},
  organization = {PMLR}
}

@article{hassanzadeh2020review,
  title     = {A review of cybersecurity incidents in the water sector},
  author    = {Hassanzadeh, Amin and Rasekh, Amin and Galelli, Stefano and Aghashahi, Mohsen and Taormina, Riccardo and Ostfeld, Avi and Banks, M Katherine},
  journal   = {Journal of Environmental Engineering},
  volume    = {146},
  number    = {5},
  pages     = {03120003},
  year      = {2020},
  publisher = {American Society of Civil Engineers}
}

@article{knoben2020brief,
  title     = {A brief analysis of conceptual model structure uncertainty using 36 models and 559 catchments},
  author    = {Knoben, Wouter JM and Freer, Jim E and Peel, MC and Fowler, KJA and Woods, Ross A},
  journal   = {Water Resources Research},
  volume    = {56},
  number    = {9},
  pages     = {e2019WR025975},
  year      = {2020},
  publisher = {Wiley Online Library}
}

@article{demargne2014science,
  title     = {The science of NOAA's operational hydrologic ensemble forecast service},
  author    = {Demargne, Julie and Wu, Limin and Regonda, Satish K and Brown, James D and Lee, Haksu and He, Minxue and Seo, Dong-Jun and Hartman, Robert and Herr, Henry D and Fresch, Mark and others},
  journal   = {Bulletin of the American Meteorological Society},
  volume    = {95},
  number    = {1},
  pages     = {79--98},
  year      = {2014},
  publisher = {American Meteorological Society}
}

@techreport{luo2015clever,
  author      = {Luo, Charles},
  title       = {Technical Reference for the {CLEVER} Model: A Real-Time Flood Forecasting Model for British Columbia},
  institution = {BC River Forecast Centre},
  address     = {Victoria, British Columbia, Canada},
  year        = {2015},
  note        = {Latest update 16 August 2017},
  url         = {https://bcrfc.env.gov.bc.ca/freshet/cleverm_ref/CLEVER_TechRef.pdf}
}

@techreport{luo2018operational,
  author      = {Luo, Charles},
  title       = {Operational Real-Time Flood Forecasting under Climate Change Impacts: The {COFFEE} Model for Coastal Storm Dominated Watersheds in British Columbia},
  institution = {BC River Forecast Centre},
  address     = {Victoria, British Columbia, Canada},
  year        = {2018},
  note        = {Revised 24 October 2019},
  url         = {https://bcrfc.env.gov.bc.ca/fallfloods/TechnicalRefCOFFEEModel.pdf}
}

@inproceedings{moore1990real,
  title     = {A real-time flow forecasting system for region-wide application},
  author    = {Moore, RJ and Jones, DA and Chadwick, AF},
  booktitle = {Urban/rural application of weather radar for flow forecasting, Proc. CEC workshop, Wageningen},
  pages     = {89--92},
  year      = {1990}
}

@article{guikema2020artificial,
  title     = {Artificial intelligence for natural hazards risk analysis: Potential, challenges, and research needs},
  author    = {Guikema, Seth},
  journal   = {Risk Analysis},
  volume    = {40},
  number    = {6},
  pages     = {1117--1123},
  year      = {2020},
  publisher = {Wiley Online Library}
}

@techreport{gao2023artificial,
  author      = {{U.S. Government Accountability Office}},
  title       = {Artificial Intelligence in Natural Hazard Modeling: Severe Storms, Hurricanes, Floods, and Wildfires},
  institution = {U.S. Government Accountability Office},
  number      = {GAO-24-106213},
  address     = {Washington, DC},
  year        = {2023},
  month       = dec,
  url         = {https://www.gao.gov/products/gao-24-106213}
}

@article{lees2021benchmarking,
  title     = {Benchmarking data-driven rainfall--runoff models in Great Britain: a comparison of long short-term memory (LSTM)-based models with four lumped conceptual models},
  author    = {Lees, Thomas and Buechel, Marcus and Anderson, Bailey and Slater, Louise and Reece, Steven and Coxon, Gemma and Dadson, Simon J},
  journal   = {Hydrology and Earth System Sciences},
  volume    = {25},
  number    = {10},
  pages     = {5517--5534},
  year      = {2021},
  publisher = {Copernicus Publications G{\"o}ttingen, Germany}
}

@article{sabzipour2023comparing,
  title     = {Comparing a long short-term memory (LSTM) neural network with a physically-based hydrological model for streamflow forecasting over a Canadian catchment},
  author    = {Sabzipour, Behmard and Arsenault, Richard and Troin, Magali and Martel, Jean-Luc and Brissette, Fran{\c{c}}ois and Brunet, Fr{\'e}d{\'e}ric and Mai, Juliane},
  journal   = {Journal of Hydrology},
  volume    = {627},
  pages     = {130380},
  year      = {2023},
  publisher = {Elsevier}
}

@article{kratzert2019toward,
  title     = {Toward improved predictions in ungauged basins: Exploiting the power of machine learning},
  author    = {Kratzert, Frederik and Klotz, Daniel and Herrnegger, Mathew and Sampson, Alden K and Hochreiter, Sepp and Nearing, Grey S},
  journal   = {Water Resources Research},
  volume    = {55},
  number    = {12},
  pages     = {11344--11354},
  year      = {2019},
  publisher = {Wiley Online Library}
}

@article{feng2023suitability,
  title   = {The suitability of differentiable, physics-informed machine learning hydrologic models for ungauged regions and climate change impact assessment},
  author  = {Feng, Dapeng and Beck, Hylke and Lawson, Kathryn and Shen, Chaopeng},
  journal = {Hydrology and Earth System Sciences},
  volume  = {27},
  number  = {12},
  pages   = {2357--2373},
  year    = {2023},
  doi     = {10.5194/hess-27-2357-2023}
}

@article{hochreiter1997long,
  title     = {Long short-term memory},
  author    = {Hochreiter, Sepp and Schmidhuber, J{\"u}rgen},
  journal   = {Neural computation},
  volume    = {9},
  number    = {8},
  pages     = {1735--1780},
  year      = {1997},
  publisher = {MIT press}
}

@article{addor2017camels,
  title     = {The CAMELS data set: catchment attributes and meteorology for large-sample studies},
  author    = {Addor, Nans and Newman, Andrew J and Mizukami, Naoki and Clark, Martyn P},
  journal   = {Hydrology and Earth System Sciences},
  volume    = {21},
  number    = {10},
  pages     = {5293--5313},
  year      = {2017},
  publisher = {Copernicus GmbH}
}

@article{zimmermann2022increasing,
  title   = {Increasing confidence in adversarial robustness evaluations},
  author  = {Zimmermann, Roland S and Brendel, Wieland and Tramer, Florian and Carlini, Nicholas},
  journal = {Advances in neural information processing systems},
  volume  = {35},
  pages   = {13174--13189},
  year    = {2022}
}

@inproceedings{wang2019improving,
  title     = {Improving adversarial robustness requires revisiting misclassified examples},
  author    = {Wang, Yisen and Zou, Difan and Yi, Jinfeng and Bailey, James and Ma, Xingjun and Gu, Quanquan},
  booktitle = {International conference on learning representations},
  year      = {2019}
}

@inproceedings{ye2019adversarial,
  title     = {Adversarial robustness vs. model compression, or both?},
  author    = {Ye, Shaokai and Xu, Kaidi and Liu, Sijia and Cheng, Hao and Lambrechts, Jan-Henrik and Zhang, Huan and Zhou, Aojun and Ma, Kaisheng and Wang, Yanzhi and Lin, Xue},
  booktitle = {Proceedings of the IEEE/CVF International Conference on Computer Vision},
  pages     = {111--120},
  year      = {2019}
}

@article{szegedy2013intriguing,
  title   = {Intriguing properties of neural networks},
  author  = {Szegedy, Christian and Zaremba, Wojciech and Sutskever, Ilya and Bruna, Joan and Erhan, Dumitru and Goodfellow, Ian and Fergus, Rob},
  journal = {arXiv preprint arXiv:1312.6199},
  year    = {2013}
}

@inproceedings{huang2011adversarial,
  title     = {Adversarial machine learning},
  author    = {Huang, Ling and Joseph, Anthony D and Nelson, Blaine and Rubinstein, Benjamin IP and Tygar, J Doug},
  booktitle = {Proceedings of the 4th ACM workshop on Security and artificial intelligence},
  pages     = {43--58},
  year      = {2011}
}

@inproceedings{dalvi2004adversarial,
  title     = {Adversarial classification},
  author    = {Dalvi, Nilesh and Domingos, Pedro and Mausam and Sanghai, Sumit and Verma, Deepak},
  booktitle = {Proceedings of the tenth ACM SIGKDD international conference on Knowledge discovery and data mining},
  pages     = {99--108},
  year      = {2004}
}

@inproceedings{barreno2006can,
  title     = {Can machine learning be secure?},
  author    = {Barreno, Marco and Nelson, Blaine and Sears, Russell and Joseph, Anthony D and Tygar, J Doug},
  booktitle = {Proceedings of the 2006 ACM Symposium on Information, computer and communications security},
  pages     = {16--25},
  year      = {2006}
}

@article{wiyatno2019adversarial,
  title   = {Adversarial examples in modern machine learning: A review},
  author  = {Wiyatno, Rey Reza and Xu, Anqi and Dia, Ousmane and De Berker, Archy},
  journal = {arXiv preprint arXiv:1911.05268},
  year    = {2019}
}

@inproceedings{papernot2017practical,
  title     = {Practical black-box attacks against machine learning},
  author    = {Papernot, Nicolas and McDaniel, Patrick and Goodfellow, Ian and Jha, Somesh and Celik, Z Berkay and Swami, Ananthram},
  booktitle = {Proceedings of the 2017 ACM on Asia conference on computer and communications security},
  pages     = {506--519},
  year      = {2017}
}

@article{goodfellow2014explaining,
  title   = {Explaining and harnessing adversarial examples},
  author  = {Goodfellow, Ian J and Shlens, Jonathon and Szegedy, Christian},
  journal = {arXiv preprint arXiv:1412.6572},
  year    = {2014}
}

@inproceedings{kurakin2017adversarial,
  title     = {Adversarial examples in the physical world},
  author    = {Kurakin, Alexey and Goodfellow, Ian J and Bengio, Samy},
  booktitle = {International Conference on Learning Representations (ICLR) Workshop},
  year      = {2017}
}

@inproceedings{mode2020adversarial,
  title        = {Adversarial examples in deep learning for multivariate time series regression},
  author       = {Mode, Gautam Raj and Hoque, Khaza Anuarul},
  booktitle    = {2020 IEEE Applied Imagery Pattern Recognition Workshop (AIPR)},
  pages        = {1--10},
  year         = {2020},
  organization = {IEEE}
}

@article{ilyas2019adversarial,
  title   = {Adversarial examples are not bugs, they are features},
  author  = {Ilyas, Andrew and Santurkar, Shibani and Tsipras, Dimitris and Engstrom, Logan and Tran, Brandon and Madry, Aleksander},
  journal = {Advances in neural information processing systems},
  volume  = {32},
  year    = {2019}
}

@inproceedings{biggio2013evasion,
  title        = {Evasion attacks against machine learning at test time},
  author       = {Biggio, Battista and Corona, Igino and Maiorca, Davide and Nelson, Blaine and {\v{S}}rndi{\'c}, Nedim and Laskov, Pavel and Giacinto, Giorgio and Roli, Fabio},
  booktitle    = {Machine Learning and Knowledge Discovery in Databases: European Conference, ECML PKDD 2013, Prague, Czech Republic, September 23-27, 2013, Proceedings, Part III 13},
  pages        = {387--402},
  year         = {2013},
  organization = {Springer}
}

@article{shah2020pitfalls,
  title   = {The pitfalls of simplicity bias in neural networks},
  author  = {Shah, Harshay and Tamuly, Kaustav and Raghunathan, Aditi and Jain, Prateek and Netrapalli, Praneeth},
  journal = {Advances in Neural Information Processing Systems},
  volume  = {33},
  pages   = {9573--9585},
  year    = {2020}
}

@article{jo2017measuring,
  title   = {Measuring the tendency of cnns to learn surface statistical regularities},
  author  = {Jo, Jason and Bengio, Yoshua},
  journal = {arXiv preprint arXiv:1711.11561},
  year    = {2017}
}

@article{geirhos2020shortcut,
  title     = {Shortcut learning in deep neural networks},
  author    = {Geirhos, Robert and Jacobsen, J{\"o}rn-Henrik and Michaelis, Claudio and Zemel, Richard and Brendel, Wieland and Bethge, Matthias and Wichmann, Felix A},
  journal   = {Nature Machine Intelligence},
  volume    = {2},
  number    = {11},
  pages     = {665--673},
  year      = {2020},
  publisher = {Nature Publishing Group UK London}
}

@article{kong2023adversarial,
  title     = {Adversarial Attacks on Regression Systems via Gradient Optimization},
  author    = {Kong, Xiangyin and Ge, Zhiqiang},
  journal   = {IEEE Transactions on Systems, Man, and Cybernetics: Systems},
  year      = {2023},
  publisher = {IEEE}
}

@article{dera2023trustworthy,
  author    = {Dera, Dimah and Ahmed, Sabeen and Bouaynaya, Nidhal C. and Rasool, Ghulam},
  title     = {{TR}ustworthy Uncertainty Propagation for Sequential Time-Series Analysis in {RNN}s},
  journal   = {IEEE Transactions on Knowledge and Data Engineering},
  volume    = {36},
  number    = {2},
  pages     = {882--896},
  month     = feb,
  year      = {2024},
  doi       = {10.1109/TKDE.2023.3288628},
  url       = {https://doi.org/10.1109/TKDE.2023.3288628},
  publisher = {IEEE}
}

@article{galib2023susceptibility,
  title   = {On the susceptibility and robustness of time series models through adversarial attack and defense},
  author  = {Galib, Asadullah Hill and Bashyal, Bidhan},
  journal = {arXiv preprint arXiv:2301.03703},
  year    = {2023}
}

@article{wang2023wasserstein,
  title   = {Wasserstein Adversarial Examples on Univariant Time Series Data},
  author  = {Wang, Wenjie and Xiong, Li and Lou, Jian},
  journal = {arXiv preprint arXiv:2303.12357},
  year    = {2023}
}

@article{ortiz2021optimism,
  title     = {Optimism in the face of adversity: Understanding and improving deep learning through adversarial robustness},
  author    = {Ortiz-Jim{\'e}nez, Guillermo and Modas, Apostolos and Moosavi-Dezfooli, Seyed-Mohsen and Frossard, Pascal},
  journal   = {Proceedings of the IEEE},
  volume    = {109},
  number    = {5},
  pages     = {635--659},
  year      = {2021},
  publisher = {IEEE}
}

@article{gan2014comprehensive,
  title     = {A comprehensive evaluation of various sensitivity analysis methods: A case study with a hydrological model},
  author    = {Gan, Yanjun and Duan, Qingyun and Gong, Wei and Tong, Charles and Sun, Yunwei and Chu, Wei and Ye, Aizhong and Miao, Chiyuan and Di, Zhenhua},
  journal   = {Environmental modelling \& software},
  volume    = {51},
  pages     = {269--285},
  year      = {2014},
  publisher = {Elsevier}
}

@article{mccuen1973role,
  title     = {The role of sensitivity analysis in hydrologic modeling},
  author    = {McCuen, Richard H},
  journal   = {Journal of hydrology},
  volume    = {18},
  number    = {1},
  pages     = {37--53},
  year      = {1973},
  publisher = {Elsevier}
}

@article{wi2024need,
  title     = {On the need for physical constraints in deep learning rainfall--runoff projections under climate change: a sensitivity analysis to warming and shifts in potential evapotranspiration},
  author    = {Wi, Sungwook and Steinschneider, Scott},
  journal   = {Hydrology and Earth System Sciences},
  volume    = {28},
  number    = {3},
  pages     = {479--503},
  year      = {2024},
  publisher = {Copernicus Publications G{\"o}ttingen, Germany}
}

@article{loritz2024camels,
  author  = {Loritz, Ralf and Dolich, Alexander and Acu{\~n}a Espinoza, Eduardo and Ebeling, Pia and Guse, Bj{\"o}rn and G{\"o}tte, Jonas and Hassler, Sibylle K. and Hauffe, Corina and Heidb{\"u}chel, Ingo and Kiesel, Jens and M{\"a}licke, Mirko and M{\"u}ller-Thomy, Hannes and St{\"o}lzle, Michael and Tarasova, Larisa},
  title   = {{CAMELS-DE}: hydro-meteorological time series and attributes for 1582 catchments in Germany},
  journal = {Earth System Science Data},
  volume  = {16},
  number  = {12},
  pages   = {5625--5642},
  year    = {2024},
  doi     = {10.5194/essd-16-5625-2024},
  url     = {https://doi.org/10.5194/essd-16-5625-2024}
}

@article{vrugt2016markov,
  title     = {Markov chain Monte Carlo simulation using the DREAM software package: Theory, concepts, and MATLAB implementation},
  author    = {Vrugt, Jasper A},
  journal   = {Environmental Modelling \& Software},
  volume    = {75},
  pages     = {273--316},
  year      = {2016},
  publisher = {Elsevier}
}

@misc{camelsde_zenodo_13837553,
  author    = {Dolich, Alexander and Espinoza, Eduardo Acu{\~n}a and Ebeling, Pia and Guse, Bj{\"o}rn and G{\"o}tte, Jonas and Hassler, Sibylle and Hauffe, Corina and Kiesel, Jens and Heidb{\"u}chel, Ingo and M{\"a}licke, Mirko and M{\"u}ller-Thomy, Hannes and St{\"o}lzle, Michael and Tarasova, Larisa and Loritz, Ralf},
  title     = {{CAMELS-DE}: hydrometeorological time series and attributes for 1582 catchments in Germany},
  year      = {2024},
  month     = sep,
  publisher = {Zenodo},
  version   = {1.0.0},
  doi       = {10.5281/zenodo.13837553},
  url       = {https://doi.org/10.5281/zenodo.13837553}
}

@article{rauthe2013central,
  title     = {A Central European precipitation climatology--Part I: Generation and validation of a high-resolution gridded daily data set (HYRAS)},
  author    = {Rauthe, Monika and Steiner, Heiko and Riediger, Ulf and Mazurkiewicz, Alex and Gratzki, Annegret and others},
  journal   = {Meteorologische Zeitschrift},
  volume    = {22},
  number    = {3},
  pages     = {235--256},
  year      = {2013},
  publisher = {Schweizerbart}
}

@article{razafimaharo2020new,
  title     = {New high-resolution gridded dataset of daily mean, minimum, and maximum temperature and relative humidity for Central Europe (HYRAS)},
  author    = {Razafimaharo, Christ{\`e}ne and Kr{\"a}henmann, Stefan and H{\"o}pp, Simona and Rauthe, Monika and Deutschl{\"a}nder, Thomas},
  journal   = {Theoretical and Applied Climatology},
  volume    = {142},
  pages     = {1531--1553},
  year      = {2020},
  publisher = {Springer}
}

@article{breuer2009assessing,
  title     = {Assessing the impact of land use change on hydrology by ensemble modeling (LUCHEM). I: Model intercomparison with current land use},
  author    = {Breuer, L and Huisman, JA and Willems, Patrick and Bormann, H and Bronstert, A and Croke, BFW and Frede, H-G and Gr{\"a}ff, T and Hubrechts, L and Jakeman, AJ and others},
  journal   = {Advances in water resources},
  volume    = {32},
  number    = {2},
  pages     = {129--146},
  year      = {2009},
  publisher = {Elsevier}
}

@article{shrestha2008data,
  title     = {Data-driven approaches for estimating uncertainty in rainfall-runoff modelling},
  author    = {Shrestha, Durga Lal and Solomatine, Dimitri P},
  journal   = {International Journal of River Basin Management},
  volume    = {6},
  number    = {2},
  pages     = {109--122},
  year      = {2008},
  publisher = {Taylor \& Francis}
}

@article{ayzel2021openforecast,
  title     = {Openforecast v2: Development and benchmarking of the first national-scale operational runoff forecasting system in russia},
  author    = {Ayzel, Georgy},
  journal   = {Hydrology},
  volume    = {8},
  number    = {1},
  pages     = {3},
  year      = {2021},
  publisher = {MDPI}
}

@article{baste2025unveiling,
  title     = {Unveiling the Limits of Deep Learning Models in Hydrological Extrapolation Tasks},
  author    = {Baste, Sanika and Klotz, Daniel and Acu{\~n}a Espinoza, Eduardo and B{\'a}rdossy, Andr{\'a}s and Loritz, Ralf},
  journal   = {Hydrology and Earth System Sciences},
  volume    = {29},
  pages     = {5871--5891},
  year      = {2025},
  doi       = {10.5194/hess-29-5871-2025},
  publisher = {Copernicus Publications}
}

@article{chahinian2006compilation,
  title   = {Compilation of the MOPEX 2004 results},
  author  = {Chahinian, Nan{\'e}e and Andr{\'e}assian, Vazken and Duan, Q and Fortin, V and Gupta, H and Hogue, T and Mathevet, T and Montanari, Andrea and Moretti, G and Moussa, Roger and others},
  journal = {IAHS publication},
  volume  = {307},
  pages   = {313},
  year    = {2006}
}

@article{cavadias1988approximate,
  title     = {Approximate confidence intervals for verification criteria of the WMO intercomparison of snowmelt runoff models},
  author    = {Cavadias, G and Morin, G},
  journal   = {Hydrological sciences journal},
  volume    = {33},
  number    = {4},
  pages     = {369--377},
  year      = {1988},
  publisher = {Taylor \& Francis}
}

@article{bardossy2022precipitation,
  title     = {Is precipitation responsible for the most hydrological model uncertainty?},
  author    = {B{\'a}rdossy, Andr{\'a}s and Kilsby, Chris and Birkinshaw, Stephen and Wang, Ning and Anwar, Faizan},
  journal   = {Frontiers in Water},
  volume    = {4},
  pages     = {836554},
  year      = {2022},
  publisher = {Frontiers Media SA}
}

@article{feng2022differentiable,
  title     = {Differentiable, learnable, regionalized process-based models with multiphysical outputs can approach state-of-the-art hydrologic prediction accuracy},
  author    = {Feng, Dapeng and Liu, Jiangtao and Lawson, Kathryn and Shen, Chaopeng},
  journal   = {Water Resources Research},
  volume    = {58},
  number    = {10},
  pages     = {e2022WR032404},
  year      = {2022},
  publisher = {Wiley Online Library}
}

@techreport{seibert2005hbv,
  author      = {Seibert, Jan},
  title       = {{HBV Light Version 2}: User's Manual},
  institution = {Department of Physical Geography and Quaternary Geology, Stockholm University},
  address     = {Stockholm, Sweden},
  year        = {2005},
  month       = nov,
  url         = {https://www.geo.uzh.ch/dam/jcr:c8afa73c-ac90-478e-a8c7-929eed7b1b62/HBV_manual_2005.pdf}
}

@article{Jansen2021HBV,
  author   = {Jansen, Koen F. and Teuling, Adriaan J. and Craig, James R. and Dal Molin, Marco and Knoben, Wouter J. M. and Parajka, Juraj and Vis, Marc and Melsen, Lieke A.},
  title    = {Mimicry of a Conceptual Hydrological Model (HBV): What's in a Name?},
  journal  = {Water Resources Research},
  volume   = {57},
  number   = {5},
  pages    = {e2020WR029143},
  doi      = {https://doi.org/10.1029/2020WR029143},
  url      = {https://agupubs.onlinelibrary.wiley.com/doi/abs/10.1029/2020WR029143},
  eprint   = {https://agupubs.onlinelibrary.wiley.com/doi/pdf/10.1029/2020WR029143},
  note     = {e2020WR029143 2020WR029143},
  year     = {2021}
}

@article{Kim2026Deep,
  author  = {Kim, Seon-Ho and Hwang, Jeongwoo and Devineni, Naresh and Sankarasubramanian, A.},
  title   = {Do Deep-Learning Models Perform Better Than Process-Based Models? A Diagnostic Evaluation Using Synthetic Streamflow Simulation Under Various Sources of Uncertainties},
  journal = {Water Resources Research},
  volume  = {62},
  number  = {5},
  pages   = {e2025WR040439},
  year    = {2026},
  doi     = {10.1029/2025WR040439}
}

@article{Courvoisier2025Assessing,
  author  = {Courvoisier, Fabien and Kraft, Basil and Haddad, Yann Yasser and Zappa, Massimiliano and Gudmundsson, Lukas},
  title   = {Assessing the Stability of {LSTM} Runoff Projections in Switzerland Under Climate Scenarios},
  journal = {EGUsphere},
  year    = {2025},
  note    = {Preprint},
  doi     = {10.5194/egusphere-2025-6058}
}

@misc{caravande_all_versions,
  author    = {Dolich, Alexander and Maharjan, Ashim and M{\"a}licke, Mirko
               and Manoj J, Ashish and Loritz, Ralf},
  title     = {Caravan-DE: Caravan Extension Germany—German Dataset
               for Large-Sample Hydrology},
  year      = {2025},
  publisher = {Zenodo},
  doi       = {10.5281/zenodo.13320514},
  url       = {https://doi.org/10.5281/zenodo.13320514}
}

@inproceedings{kireev2022effectiveness,
  title     = {On the Effectiveness of Adversarial Training Against Common Corruptions},
  author    = {Kireev, Klim and Andriushchenko, Maksym and Flammarion, Nicolas},
  booktitle = {Proceedings of the Thirty-Eighth Conference on Uncertainty in Artificial Intelligence},
  series    = {Proceedings of Machine Learning Research},
  volume    = {180},
  pages     = {1012--1021},
  year      = {2022},
  publisher = {PMLR},
  url       = {https://proceedings.mlr.press/v180/kireev22a.html}
}

  \end{document}